\documentclass[12pt]{article}
\usepackage[OT1]{fontenc}
\usepackage{amsthm,amsmath,amssymb,amsfonts,mathrsfs,amsfonts,dsfont}
\usepackage{bm,enumerate,xcolor,animate,graphicx,graphics,natbib,color,url}
\usepackage[colorlinks,citecolor=blue,urlcolor=blue]{hyperref}
\usepackage{booktabs,float,lscape} 
\usepackage{mdframed,verbatim}
\usepackage{subfig,pifont}
\usepackage[absolute]{textpos}
\allowdisplaybreaks

\def\black#1{{\color{black} #1}}%
\def\pr{\text{Pr}}%

\def\E{\text{E}}

\def\ss{\mathbf{s}}

\numberwithin{equation}{section}
\theoremstyle{plain}

\theoremstyle{remark}

\theoremstyle{remark}

\newcommand{\blind}{1}

\graphicspath{{/Users/castroda/Dropbox/Projects/ExtremeWindForecasting/Paper/arXiv/Version2/}}

\begin{document}
\thispagestyle{empty}
\baselineskip=28pt

%-------------------------------------------------------------------------------
%	TITLE, AUTHORS INFORMATION
%-------------------------------------------------------------------------------
\vskip 5mm
\begin{center} {\Large{\bf A  {spliced Gamma-Generalized Pareto} model for short-term extreme wind speed probabilistic forecasting}}
\end{center}

\baselineskip=12pt
\vskip 5mm
\if1\blind
{
\begin{center}
\large
Daniela Castro-Camilo$^1$, Rapha\"el Huser$^2$, and H{\aa}vard Rue$^2$
\end{center}
\footnotetext[1]{
\baselineskip=10pt School of Mathematics and Statistics, University of Glasgow, Glasgow G12 8QQ, UK. E-mail: daniela.castro.camilo@gmail.com.}
\footnotetext[2]{
\baselineskip=10pt Computer, Electrical and Mathematical Sciences and Engineering (CEMSE) Division, King Abdullah University of Science and Technology (KAUST), Thuwal 23955-6900, Saudi Arabia. E-mails: raphael.huser@kaust.edu.sa, haavard.rue@kaust.edu.sa.}

} \fi

\baselineskip=17pt
\vskip 4mm
\centerline{June 17, 2019}
\vskip 6mm

%-------------------------------------------------------------------------------
%	ABSTRACT
%-------------------------------------------------------------------------------
\begin{center}
{\large{\bf Abstract}}
\end{center}
Renewable sources of energy such as wind power have become a sustainable alternative to fossil fuel-based energy. However, the uncertainty and fluctuation of the wind speed derived from its intermittent nature bring a great threat to the wind power production stability, and to the wind turbines themselves. Lately, much work has been done on developing models to forecast average wind speed values, yet surprisingly little has focused on proposing models to accurately forecast extreme wind speeds, which can damage the turbines. In this work, we develop a flexible  {spliced Gamma-Generalized Pareto} model to forecast extreme and non-extreme wind speeds simultaneously. Our model belongs to the class of latent Gaussian models, for which {inference is conveniently performed based on} the integrated nested Laplace approximation method. Considering a flexible additive regression structure, we propose two models for the latent linear predictor to capture the spatio-temporal dynamics of wind speeds. 
%The first linear predictor is a temporal model that incorporates spatial information  {through} lagged off-site predictors, chosen as a function of the dominant wind directions. The second one uses stochastic partial differential  {equation} approximations to the Mat\'ern covariance of a Gaussian field that varies in time according to a first-order autoregressive process. 
Our models are fast to fit and can describe both the bulk and the tail of the wind speed distribution while producing short-term extreme and non-extreme wind speed probabilistic forecasts.

\baselineskip=16pt
\par\vfill\noindent

\baselineskip=26pt

%-------------------------------------------------------------------------------
%	INTRODUCTION
%-------------------------------------------------------------------------------

%-------------------------------------------------------------------------------
% SUBSECTION
%-------------------------------------------------------------------------------
\section{Introduction}\label{sec:Introduction}
 {The integration of renewable energies such a wind power into power systems has been developing on a large scale around the world in the last decade. Wind power generation is sustainable, emission-free, and its cost is nearly the same as that of coal or nuclear energy~\citep{hering2010powering}. These advantages are counter-balanced by several challenges, such as high variability, limited dispatchability, and non-storability~\citep{pinson2012adaptive,hering2010powering}. 
 {Accurate short-term forecasts of wind power are therefore {crucial} for power production planning and risk assessment.}
}

%Wind power production data observed at fine temporal resolutions exhibit successive periods with fluctuations of various dynamic nature and magnitude~\citep{pinson2012adaptive}. This high variability makes the wind power a non-steady, non-constant source of energy. 
%Moreover, unlike traditional sources of energy, wind power is not fully dispatchable. Wind farms cannot increase their power generation upon request when there is not sufficient wind; they can only reduce the output. Furthermore, wind cannot be stored for future power generation, and wind-powered energy should be used as it becomes available. 

\begin{figure}[t!]
\centering
\includegraphics[width=12cm, height=6.5cm]{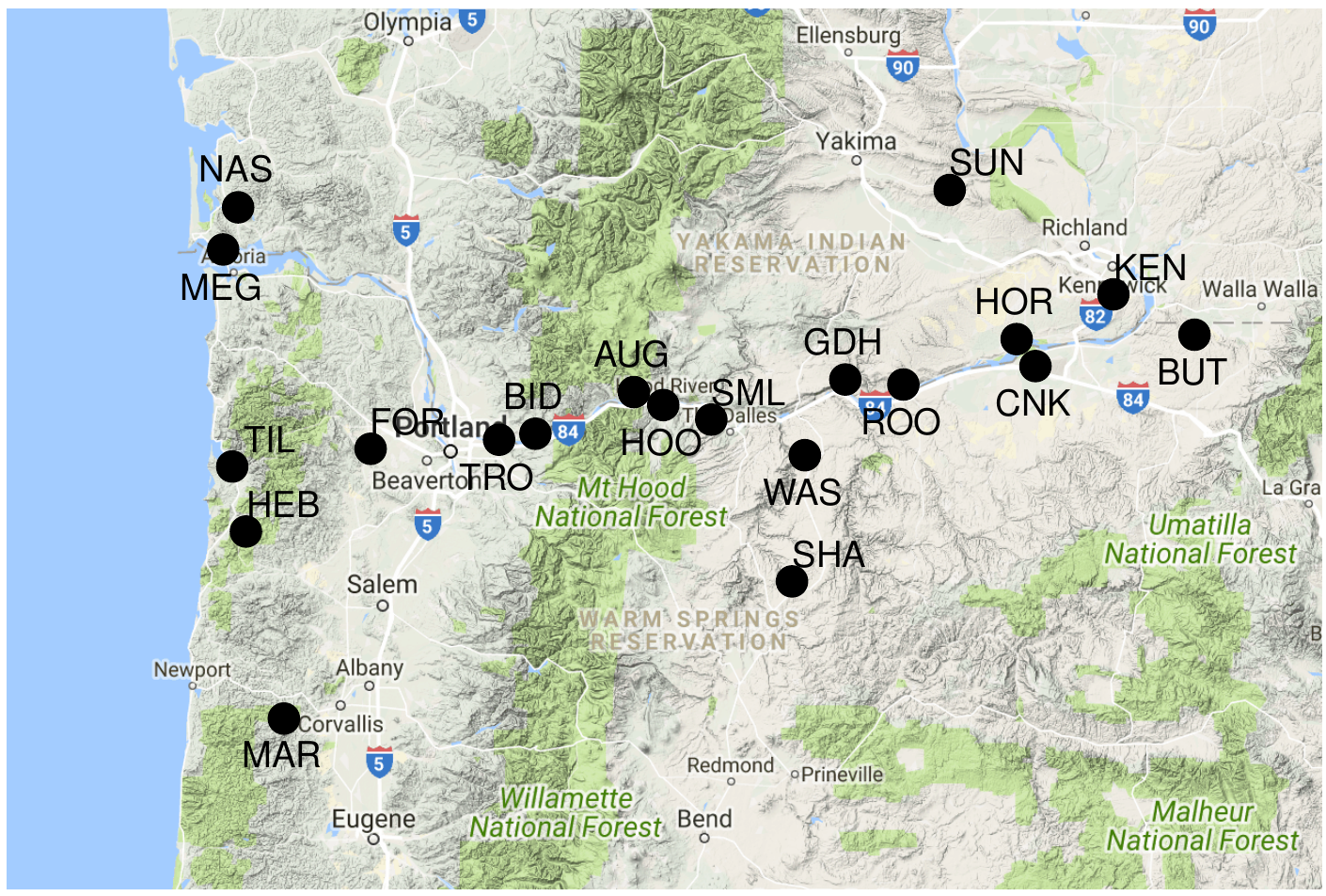}	
\caption{\footnotesize Map showing the 20 towers along the Columbia River.}
\label{fig:map.pdf}
\end{figure}

Wind power forecasts can be made directly if power data are available~\citep{lenzi2017spatial}. However, wind speed forecasting can be more precise than wind power forecasting due to the spatial correlation of wind,  {and wind power forecasts can be easily obtained from the forecasted wind speeds (see, e.g., the deterministic cubic power curve in Figure 5 in \citeauthor{hering2010powering}, \citeyear{hering2010powering})}. In this paper, we focus on wind speed probabilistic forecasting based on  {20 turbine towers measuring hourly average wind speed and wind direction. The towers are} installed at the border between Oregon and Washington, along the Columbia River (see Figure~\ref{fig:map.pdf}).
 {Each station encompasses between $T=21,306$ and $T=26,304$ hourly measurements of non-zero wind speed from January 2012 to December 2014. Basic exploratory analyses unveil different wind regimes among stations, high autocorrelation within stations, {seasonal} patterns, and persistence, which refers to the variable's tendency to maintain its current state. See Section 2 of the Supplementary Material for accompanying graphical results.}

Over the last decades, statistical models have shown to be very effective in capturing the fluctuating characteristics of wind speed to produce accurate short-term wind speed forecasts (see, e.g.,~\citeauthor{zhu2012short}, \citeyear{zhu2012short}). {Purely temporal models are built assuming that wind speed at each time point is partially predicted by wind speed in its near past. In this context, different time series models have been proposed for short-term forecasts, such as autoregressive models~\citep{huang1995use}, autoregressive moving average models~\citep{erdem2011arma}, and autoregressive integrated moving average models~\citep{palomares2009arima}. Wind speed is modeled using neural networks in~\cite{li2010comparing}, while a model to assess wind potential using spectral analysis is proposed in~\cite{shih2008wind}. Models that incorporate both temporal and spatial correlations in the form of off-site information {(i.e., information that is not collected at the site of interest, but at neighboring sites)} have been found to increase accuracy over conventional time series models. In this framework, \cite{alexiadis1999wind} use off-site predictors to improve wind speed and wind power forecasts, while~\cite{gneiting2006calibrated}, \cite{hering2010powering}, and \cite{kazor2015role} assume that wind speeds follow a truncated normal distribution with regime-dependent mean and variance. See~\cite{zhu2012short} for a review on statistical wind speed forecasting models.}

Our main goal is to develop a flexible  {space-time model} {designed} to forecast future wind speeds at the turbine locations, with a particular focus on extreme wind speeds. 
The statistical modeling of extremes is challenging, as it usually focuses on the estimation of high (or low) quantiles, where the available data are limited.  {To tackle this, we {rely on} extreme-value theory, which offers a rigorous mathematical framework to develop techniques and models for describing the tail of the distribution}; see \cite{davison2015statistics} for a review.

In the context of wind-powered energy, {a prolonged period of large wind speed values may pose} a considerable risk to the wind turbines, thus affecting wind energy production. {In this context,} a suitable definition of extremes is given by the \emph{wind speed exceedances}, i.e., wind speeds that exceed a certain high threshold $u>0$. The stochastic behavior of exceedances is described by the conditional probability $\pr(Y > u+y \mid Y>u) = \{1-F(u+y)\}/\{1-F(u)\}$, $y>0,$
where $Y$ has distribution $F$. Under broad conditions~\citep{davison1990models}, it can be shown that as $u$ becomes large, this conditional probability can be approximated by $1 - H_{GP}(y) = (1 + \xi y/\sigma_u)_+^{-1/\xi}$, where $H_{GP}$ denotes the generalized Pareto (GP) distribution function with scale parameter $\sigma_u>0$, shape parameter $\xi\in\mathbb{R}$, and $a_+ = \max(0,a)$. As $\xi\to0$, the GP distribution corresponds to the exponential distribution function $1-\exp(-y/\sigma_u)$, $y>0$. The shape parameter determines the weight of the upper tail: {a} heavy tail {is} obtained with $\xi >0$, {a} light tail with $\xi \to0$, and {a} bounded tail with $\xi < 0$. 

 {In this paper, we propose a hierarchical Bayesian {model based on the} spliced Gamma-Generalized Pareto (Gamma-GP) {distribution} to describe both non-extreme and extreme wind speeds in space and time. {Our model is} designed to provide upper tail forecasts beyond observed wind speed values. Hierarchical Bayesian models for threshold exceedances were introduced by~\cite{casson1999spatial} and \cite{cooley2007bayesian} and fitted using expensive Markov chain Monte Carlo (MCMC) methods. Spliced extreme-value and alternative {extended GP} models were {proposed} by various authors~\citep{tancredi2006accounting,naveau2016modeling} to model extreme and non-extreme data simultaneously; see, \cite{scarrott2012review} for a review and~\cite{opitz2018inla} for a recent related contribution. We here explore two different latent processes {that drive} the space-time trends and dependence structure.} {The first one is a temporal model that incorporates spatial information in the form of off-site predictors, while the second one is a space-time model.} As our {Bayesian} approach {relies} on latent Gaussian {processes}, we can exploit the integrated nested Laplace approximation (INLA;~\citeauthor{rue2009approximate}, \citeyear{rue2009approximate}) {of posterior distributions,} conveniently implemented in the R-INLA package, which provides fast and accurate inference~\citep{rue2009approximate} {for this class of models}.

The remainder of the article is organized as follows. In Section~\ref{sec:modelingextremes} we develop our modeling strategy, while in Section~\ref{sec:inference} we explain our Bayesian estimation approach using INLA. Section~\ref{sec:speedresults} presents the results of our modeling approaches in terms of wind speed forecasting. Conclusions {and an outlook towards future research} are given in Section~\ref{sec:conclusion}.

%-------------------------------------------------------------------------------
%	MODELING WIND SPEEDS
%-------------------------------------------------------------------------------
\section {Modeling wind speeds using latent Gaussian models}\label{sec:modelingextremes}

%-------------------------------------------------------------------------------
% SUBSECTION
%-------------------------------------------------------------------------------
\subsection{General framework}\label{subsec:generalframework}
Latent Gaussian models are a broad and flexible class of models, which are well-suited for modeling (possibly non-Gaussian and non-stationary) spatio-temporal data~\citep{rue2009approximate}. They admit a hierarchical model formulation, whereby observations $\mathbf{y}$ (with components $y(i), i\in\mathcal{I}$, {for some index set $\mathcal{I}$}) are {typically} assumed to be conditionally independent given a latent Gaussian random field $\mathbf{x}$ (with components $x(i), i\in\mathcal{I}$) and hyperparameters $\boldsymbol{\theta}_1$, i.e.,
\begin{align}\label{eq:hierarchicalrep}
	\mathbf{y} \mid \mathbf{x},\boldsymbol{\theta}_1 &\sim \prod_{i\in\mathcal{I}}\text{p}(y(i) \mid {x}(i),\boldsymbol{\theta}_1),\nonumber\\
	\mathbf{x} \mid \boldsymbol{\theta}_2 &\sim \mathcal{N}(\boldsymbol{\mu}_{\boldsymbol{\theta}_2}, \boldsymbol{Q}^{-1}_{\boldsymbol{\theta}_2}),\nonumber\\
	\boldsymbol{\theta} = (\boldsymbol{\theta}_1^T, \boldsymbol{\theta}_2^T)^T&\sim \text{p}(\boldsymbol{\theta}),
\end{align}
{where $\text{p}$ denotes a generic distribution.} The latent Gaussian random field $\mathbf{x}$  {has mean vector $\boldsymbol{\mu}_{\boldsymbol{\theta}_2}$ and precision matrix $\boldsymbol{Q}_{\boldsymbol{\theta}_2}$, which are controlled by the hyperparameters $\boldsymbol{\theta}_2$}. It describes the {trends and the} underlying dependence structure of the data, and its specification in~\eqref{eq:hierarchicalrep} is key to generate a flexible and versatile class of models. We {here} assume that $y(i)$ only depends on a linear predictor $\eta(i)$ which has an additive structure with respect to some fixed covariates and random effects, i.e.,
\begin{equation}\label{eq:linearpredictor}
\eta(i) = \mu + \sum_{j=1}^J\beta_jz_{j}(i) + \sum_{k=1}^Kf_{k}(w_{k}(i)),\quad {i\in\mathcal{I}}.
\end{equation}
Here, $\mu$ is the overall intercept, $z_{j}(i)$ are known covariates with coefficients $\boldsymbol{\beta} = (\beta_1,\ldots, \beta_J)^T$, and $\mathbf{f}=\{f_1(\cdot),\ldots,f_K(\cdot)\}$ are specific (a priori independent) Gaussian processes  defined in terms of a set of covariates $\mathbf{w}=(w_1(i),\ldots,w_K(i))^T$. If we further assume that $(\mu,\boldsymbol{\beta}^T)^T$ has a Gaussian prior, 
%then {we can obtain the original formulation in~\eqref{eq:hierarchicalrep}}; see \cite{rue2017bayesian} for more details.
the joint distribution of $\mathbf{x}= (\boldsymbol{\eta}^T,\mu,\boldsymbol{\beta}^T,\mathbf{f}^T)^T$ is Gaussian, and yields the latent field $\mathbf{x}$ in the hierarchical formulation~\eqref{eq:hierarchicalrep}; see e.g., \cite{rue2017bayesian} for more details. 
 {Note that in Section~\ref{subsec:offsitemodel} the index set $\mathcal{I}$ denotes time, while in Section~\ref{subsec:spdemodel}, it denotes space and time.}

%To construct a suitable linear predictor for our wind speed forecasting problem, we must take into account the properties detailed in Section~\ref{sec:Introduction}, i.e., space-time dependence, seasonality, and persistence. In the following sections, we propose two different models for the linear predictors to capture these characteristics. 

{The remainder of this section is organized following the hierarchical representation in~\eqref{eq:hierarchicalrep}. {Therefore,} Section~\ref{subsec:splicedgammagp} presents our model likelihood (with vector of parameters $\boldsymbol{\theta}_1$), Section~\ref{subsec:latentstructure} introduces our two proposed latent structures, called the off-site model and the SPDE model (with vector of parameters $\boldsymbol{\theta}_{2, \text{off-site}}$ and $\boldsymbol{\theta}_{2, \text{SPDE}}$, respectively), and Section~\ref{subsec:priors} {specifies} prior distributions for all the parameters involved, namely $\boldsymbol{\theta}_1$, $\boldsymbol{\theta}_{2, \text{off-site}}$, and $\boldsymbol{\theta}_{2, \text{SPDE}}$.}

%-------------------------------------------------------------------------------
% SUBSECTION
%-------------------------------------------------------------------------------
\subsection{A spliced Gamma-GP model}\label{subsec:splicedgammagp}

 {Here we describe the top layer of the hierarchical representation in~\eqref{eq:hierarchicalrep}, namely, the model likelihood. Loosely speaking, our model assumes that {wind speeds} under a certain threshold follow a Gamma distribution, while those above the threshold follow a generalized Pareto (GP) distribution. The (non-stationarity) threshold is estimated first using Gamma quantile regression, and it is then used subsequently to estimate the proportion of observations above the threshold using a Bernoulli distribution, as well as to fit the GP distribution. For ease of exposition, we describe our model likelihood in three stages corresponding to the Gamma, Bernoulli, and GP models, respectively.}

{\bf Stage 1.} We assume that positive wind speeds {$y(i)>0$} are characterized by a Gamma distribution parametrized in terms of an $\alpha$-quantile  {$\psi_{\alpha}(i)>0$ and a precision parameter $\kappa>0$. With this parametrization, the Gamma density is}
\begin{equation*}\label{eq:gammadensity}
h_{\text{Ga}}(y) = \frac{1}{\Gamma(\kappa)}\left\{\frac{H_{\text{Ga}}^{-1}(\alpha; \kappa, 1)}{\psi_{\alpha}(i)}\right\}^{\kappa} y^{\kappa - 1}\exp\left\{-y\frac{H_{\text{Ga}}^{-1}(\alpha; \kappa, 1)}{\psi_{\alpha}(i)}\right\},
\end{equation*}
where $H_{\text{Ga}}^{-1}(\alpha; \kappa, 1)$ is the quantile function of the Gamma distribution with unit rate and shape $\kappa$, evaluated at the probability $\alpha\in(0,1)$.
 
 {\bf Stage 2.} The value $\alpha$ corresponds to the probability that wind speeds exceed the threshold $\psi_{\alpha}(i)$, conditional on being positive. Here we estimate the unconditional probability of exceeding $\psi_{\alpha}(i)$ to account for the few zero  {wind speed} values that were previously excluded {in Stage 1}.  {Writing $y(i)$ as the {$i$-th} wind speed observation,} we model exceedance indicators $ \mathds{1}\{y(i) > \psi_{\alpha}(i)\}$ using the Bernoulli distribution $h_{\text{B}}(y) = \text{p}(i)^{y}\{1-\text{p}(i)\}^{1-y}$, $y\in\{0,1\},$ where $\text{p}(i)=\pr\{y(i) > \psi_{\alpha}(i)\}$, and $\psi_{\alpha}(i)$ has been previously estimated in Stage 1. 

 {\bf Stage 3.} Since the tail of the Gamma distribution decays exponentially fast, probabilities associated with extreme events might be underestimated  {if the true underlying distribution is heavy-tailed. To more flexibly model} the probabilities associated with extreme events, we correct the tail using a GP distribution; recall Section~\ref{sec:Introduction}. Threshold exceedances defined as  {$x(i) = \{y(i) - \psi_{\alpha}(i)\}\mid y(i) > \psi_{\alpha}(i)$, are characterized by a GP distribution parametrized in terms of a $\beta$-quantile $\phi_{\beta}(i)>0$ and a shape parameter $\xi\geq 0$.} With this parametrization, the GP distribution is
\begin{equation*}\label{eq:q-gpd}
H_\text{GP}(y) = 
     \begin{cases}
       1 - \left[1+\{(1-\beta)^{-\xi}-1\}y/\phi_{\beta}(i)\right]^{-1/\xi}, &\quad\text{if }\xi > 0,\vspace{0.3cm}\\
       1 - (1-\beta)^{y/\phi_{\beta}(i)}, &\quad\text{if }\xi = 0, \\
     \end{cases}\quad y>0.
\end{equation*}

Note that the goal of the Gamma  {distribution in Stage 1} is twofold: to describe the distribution of non-extreme wind speeds and to obtain a suitable threshold $\psi_{\alpha}(i)$ to define wind speed exceedances. The second and third stages model the frequency and intensity of extremes, respectively, and are connected to the first stage through the threshold $\psi_{\alpha}(i)$.

%-------------------------------------------------------------------------------
% SUBSECTION
%-------------------------------------------------------------------------------
\subsection{Latent structure}\label{subsec:latentstructure}
%-------------------------------------------------------------------------------

 Here we detail two approaches to describe the middle layer of the hierarchical representation in~\eqref{eq:hierarchicalrep}, namely, the latent Gaussian process specification. {The first approach, called the off-site latent model, corresponds to a {temporal model fitted at each station separately,} with off-site information included in the form of covariates. In this case, $\mathcal{I} \equiv \mathcal{T} = \{1,\ldots,T\}$ ({with $T$} the number of time points) and we denote by $y_\ss(t)$ the wind speed at {some fixed} location $\ss\in\mathcal{S}\subset\mathbb{R}^2$ and time $t\in\mathcal{T}$. The second approach, called the SPDE latent  model, corresponds to a {proper} space-time {model} {stemming} from a {particular} stochastic partial differential equation (SPDE). In this case, $\mathcal{I} \equiv \mathcal{S}\times\mathcal{T}$ and we denote by $y(\ss,t)$ the wind speed at location $\ss\in\mathcal{S}$ and time $t\in\mathcal{T}$.}
% SUBSUBSECTION
%-------------------------------------------------------------------------------
\subsubsection{Off-site latent model}\label{subsec:offsitemodel}
 {For each fixed location $\ss\in\mathcal{S}$, we propose the following linear predictor:}
\begin{equation}\label{eq:our1linearpredictor}
\eta_\ss^{(1)}(t)=\mu_{\ss} + \sum_{j=1}^{|N_\ss|}\beta_{j}y_{\ss_j}(t-1) + f_{1}(t;\rho_{\ss,1},\tau_{\ss,1}) + f_{2}(w_2(t);\tau_{\ss,2}),\quad t=1,\ldots,T,
\end{equation}
where for each $j\in\{1,\ldots, |N_\ss|\}$, $y_{\ss_j}(t-1)$ is the lagged time series of wind speeds at the $j$-th neighbor of $\ss$, $N_\ss$ is the set of neighbors of $\ss$ of cardinality $|N_\ss|$. The coefficients $\{\beta_{j}\}$ (fixed effects) quantify the effect that wind speeds at the $j$-th neighbor observed at time lag one have on the response. The random effects ${f}_{1}(t) \equiv f_{1}(t;\rho_{\ss,1},\tau_{\ss,1})$ and ${f}_2(w(t)) \equiv f_{2}(w_2(t);\tau_{\ss,2})$ account for unobserved heterogeneity within each station. Specifically, ${f}_{1}(t)$  {captures the autocorrelation structure of order one within each wind speed time series, and is assumed to be described by a} zero-mean autoregressive Gaussian process of first order, i.e.,
		\begin{eqnarray*}
		{f}_{1}(1)&\sim&\mathcal{N}(0,\{\tau_{\ss,1}(1-\rho_{\ss,1}^2)\}^{-1}),\\
		{f}_{1}(t)&=&\rho_{\ss,1} {f}_{1}(t-1) + \epsilon(t), \quad |\rho_{\ss,1}|<1,\quad \epsilon(t)\sim\mathcal{N}(0, \tau_{\ss,1}^{-1}),\quad t= 2,\ldots,T,
	\end{eqnarray*}
	where $\tau_{\ss,1}>0$ is a precision parameter.  {Our experiments show that including greater lags do not significantly improve the fit or predictive power of our model. The random effect ${f}_{1}(t)$ along with the lagged covariates account for temporal dependence and persistence.}

	 {The random effect ${f}_2(w(t))$ represents the sub-daily variation of wind speeds,  {and is similar in spirit to the semiparametric splines of~\cite{hering2015markov}}. It} is assumed to be described by a cyclic Gaussian random walk of second order with precision $\tau_{\ss,2}>0$, defined over each of the 24 hours within a day~\citep[Ch. 3]{rue2005gaussian}. Let $w_2(t)\in\{1,\ldots,24\}$ denote the hour associated to time $t$, then:
	\begin{equation*}
		f_2(w_2(t))-2f_2(w_2(t+1))+f_2(w_2(t+2))\sim\mathcal{N}(0, \tau_{\ss,2}^{-1}),\quad t=1,\ldots, T-2.
	\end{equation*}

Both random effects are constrained to sum to zero, and their precision hyperparameters $\tau_{\ss,l}$, $l=1,2$, have the main goal of controlling the strength of dependence among neighboring covariate classes, {i.e., the smoothness of the corresponding random effect.}

{The parameters of our spliced Gamma-GP model are linked to different linear predictors of the form~\eqref{eq:our1linearpredictor}. For consistency with the off-site model notation, we now write $\psi_{\ss,\alpha}(t)$, $\text{p}_\ss(t)$, and $\phi_{\ss,\beta}(t)$ to indicate time-varying but location-specific Gamma $\alpha$-quantile, Bernoulli probability, and GP $\beta$-quantile, respectively, and $\kappa_\ss$ and $\xi_\ss$ to indicate {time-constant, but} location-specific Gamma and GP shape parameters, respectively. The time-varying parameters are {finally} linked to the off-site latent model as follows:}
\begin{align}\label{eq:link}
	\text{Stage 1:}\qquad&\psi_{\ss,\alpha}(t)=\exp\{\eta_{\ss,\text{Gamma}}^{(1)}(t)\},\quad\text{where}\quad \pr\{y_\ss(t)\leq \psi_{\ss,\alpha}(t)\} = \alpha,\nonumber\\
	\text{Stage 2:}\qquad&\text{p}_\ss(t)=\exp\{\eta_{\ss, \text{Ber}}^{(1)}(t)\}/[1 + \exp\{\eta_{\ss,\text{Ber}}^{(1)}(t)\}],\nonumber\\
	\text{Stage 3:}\qquad&\phi_{\ss,\beta}(t) =\exp\{\eta_{\ss, \text{GP}}^{(1)}(t)\},\quad \text{where}\quad\pr\{y_\ss(t)\leq \phi_{\ss,\beta}(t)\} = \beta. 
\end{align}

%-------------------------------------------------------------------------------
% SUBSUBSECTION
%-------------------------------------------------------------------------------
\subsubsection{SPDE latent model}\label{subsec:spdemodel}
The linear predictor in~\eqref{eq:our1linearpredictor} corresponds to a temporal model that introduces spatial information using lagged off-site predictors. Here we {describe} a proper space-time model, which explicitly accounts for spatial dependence amongst wind speeds at different stations. Specifically, we assume that the space-time dependence between wind speeds at different wind towers can be described by a spatio-temporal term $u(\ss,t)$ that varies in time according to a first-order autoregressive structure. Specifically, we assume that $u(\ss,t) = \rho_2u(\ss,t-1) + z(\ss,t),$
%\begin{equation*}
%	u(\ss,t) = \rho_2u(\ss,t-1) + z(\ss,t),
%\end{equation*}
where $|\rho_2|<1$, and $z(\ss,t)$ is a zero-mean, temporally independent Gaussian field, that is completely determined by a stationary M\'atern covariance function with marginal variance $\sigma^2>0$, range $r>0$, and fixed smoothness parameter. 
%its covariance function   
%\begin{equation*}
%	\text{Cov}\{z(\ss,t), z(\ss',t')\} = \begin{cases} 
%      \sigma^2C(h;r), & \text{if } t = t', \\
%      0, & \text{if } t \neq t', \\
%   \end{cases}\quad\text{} h = ||\ss- \ss'||,
%\end{equation*}
%where  {$\sigma^2 >0$ is the marginal variance} and $C(h;r)$ is the stationary M\'atern correlation function (see, e.g., \citeauthor{handcock1993bayesian}, \citeyear{handcock1993bayesian})  {with range $r>0$ and fixed smoothness parameter}. 
This gives rise to our second linear predictor:
\begin{equation}\label{eq:our2linearpredictor}
\eta^{(2)}(\ss, t)= \mu + u(\ss,t) +  f_{2}(w_2(t);\tau_2),\quad \ss\in\mathcal{S},\text{ } t=1,\ldots,T,
\end{equation}
where $\mu$ is an intercept, $f_2(w_2(t)) \equiv f_{2}(w_2(t); \tau_2)$ is the cyclic random effect described in Section~\ref{subsec:offsitemodel}, capturing the sub-daily variations that are common to all the stations, and $\tau_2>0$ is its precision parameter.  {By contrast with~\eqref{eq:our1linearpredictor}, the notation $\eta^{(2)}(\ss,t)$ in~\eqref{eq:our2linearpredictor} emphasizes that this second linear predictor is a function of space and time.}

The latent process  {associated with}~\eqref{eq:our2linearpredictor} has dense covariance and (possibly) precision matrices, which implies that any attempt to make Bayesian inference can be computationally demanding. To avoid this issue, we use the Stochastic Partial Differential Equations (SPDE) approach introduced by~\cite{lindgren2011explicit}. The SPDE approach consists {in} constructing a continuous approximation to the Gaussian field $z(\ss,t)$ by using a continuous SPDE latent model defined {over} the entire study area. It can be shown that this SPDE has a Gaussian field with a Mat\'ern covariance function as stationary solution. Under certain conditions (see, e.g., \citeauthor{lindgren2011explicit}, \citeyear{lindgren2011explicit}), the continuous SPDE solution has the Markovian property. This property produces sparse precision matrices that can be easily factorized, and are the focus of the INLA methodology. An approximate discrete solution of the SPDE in a bounded domain, defined in our case by the location of the wind towers, {can be} obtained using a finite element method that allows for flexible boundaries and different levels of accuracy for the discretization; see Section 3 of the Supplementary Material for details on the mesh used to discretize the study region. For more details on spatial modeling using the SPDE approach, see the recent review by~\cite{bakka2018spatial}.

Because the Cascade Mountains illustrated in Figure~\ref{fig:map.pdf} naturally divide the study region {into} two sub-regions, we consider two {independent} sub-models for $u(\ss,t)$, one for stations to the East and another for stations to the West of the mountains. We fit each set of stations separately but using the same latent structure specified in~\eqref{eq:our2linearpredictor}.

The link between this new linear predictor and the model likelihood in Section~\ref{subsec:splicedgammagp} is the same as in~\eqref{eq:link}, {with the suitable change of notation}. {Note that by pooling stations together, the {space-time} parameters $\psi_{\alpha}(\ss, t)$ and $\phi_{\beta}(\ss, t)$ vary in space and time, whereas the Gamma and GP shape parameters {$\kappa$ and $\xi$, {respectively}}, are now fixed across locations to limit the number of hyperparameters to be estimated using INLA.}

%-------------------------------------------------------------------------------
% SUBSECTION
%-------------------------------------------------------------------------------
\subsection{Prior specification}\label{subsec:priors}
% Priors for hyperparameters in the likelihood
% Priors for hyperparameter in the off-site latent model
% Priors for hyperparameter in the SPDE latent model
 {Here we describe the bottom layer of the hierarchical representation in~\eqref{eq:hierarchicalrep}, namely, the specification of prior distributions. We need to specify priors for the likelihood hyperparameters $\boldsymbol{\theta}_1$ {($\boldsymbol{\theta}_1 = \kappa$ for the Gamma model and $\boldsymbol{\theta}_1 = \xi$ for the GP model; recall that $\kappa = \kappa_\ss$ and $\xi = \xi_\ss$ for the off-site model)}, as well as for the hyperparameters of the two latent structures described in Section~\ref{subsec:latentstructure}, namely $\boldsymbol{\theta}_{2, \text{off-site}} = (\rho_{\ss,1}, \tau_{\ss,1}, \tau_{\ss,2})^T$ and $\boldsymbol{\theta}_{2, \text{SPDE}} = (\sigma^2, r, \rho_{2}, \tau_{2})^T$.}

When little expert knowledge is available, a common practice is to assume non-informative priors. Alternatively, informative priors can be proposed using Penalized Complexity (PC) priors~\citep{simpson2017penalising}. In this framework, model components are considered to be flexible extensions of simpler base models. Priors are then developed in such a way that the components shrink towards their base models, thus preventing overfitting.~\cite{simpson2017penalising} propose to use the Kullback-Leibler divergence to measure the squared ``distance'' from the base model to its flexible extension, and to penalize this ``distance'' at constant rate.

For our likelihood hyperparameters, we assume a slightly informative prior over the Gamma shape {$\kappa$}, by considering a Gamma distribution with shape 10 and rate 1, which gives a high probability to values between 5 and 15. A {moderately} strong PC prior is assumed for the shape parameter of the GP distribution {$\xi$}; since large values of the shape parameter are usually unrealistic for wind speeds, we here assume that $\pr({\xi} > 0.4) \approx 0.01$.
 
 {Regarding the hyperparameters of the off-site latent model}, we assume fairly informative PC priors for the correlation hyperparameter of the AR(1) process, and the precision hyperparameter of the random walk of order 2. Specifically, $\pr(\rho_{\ss,1} > 0.9) = 0.95$ and $\pr(1/\sqrt{\tau_{\ss,2}} > \text{sd}_{\text{wind}}) = 0.01$, where $\text{sd}_{\text{wind}}$ denotes the empirical standard deviation of the temporally aggregated wind speeds. A diffuse prior is assumed for $\tau_{\ss,1}$, the precision hyperparameter of the AR(1) component.

 {PC priors on the parameters of the Gaussian field in the SPDE latent model, namely the marginal variance $\sigma^2>0$ and the range of dependence $r>0$}, are chosen such that the variance is shrunk towards zero, whereas the range is shrunk towards infinity~\citep{fuglstad2018constructing}. Specifically, we set $\pr(\sigma > 2\times \text{sd}_{\text{wind}})= 0.01$ and $\pr(r < r_{\text{median}}) = 0.5$, where $r_{\text{median}}$ is the median of the distances between stations. For stations to the East of the Cascade Mountains, $r_{\text{median}} = 94.6$ km, and for stations to the West, $r_{\text{median}} = 113.3$ km. A PC prior is also chosen for the correlation coefficient of the autoregressive term in $u(\ss,t)$, specifically $\pr(\rho_2 > 0.9) = 0.95$.  {The PC prior for $\tau_{2}$ is the same as for $\tau_{\ss,2}$ in the off-site latent model.} 
 
 {In general, {our} prior {specification tries} to reflect some characteristics observed in the data. As part of the model selection, we conducted a small sensitivity analysis using a subset of the data. The results show that the prior {specification} does not affect the results considerably.}
%-------------------------------------------------------------------------------
%	INFERENCE
%-------------------------------------------------------------------------------
\section {Inference based on INLA}\label{sec:inference}
Here we describe the form of the joint posterior distribution for each stage of our spliced Gamma-GP model detailed in Section~\ref{sec:modelingextremes}. In the following, {we reuse the generic notation in Section~\ref{subsec:generalframework}} for {ease of exposition}. Let $\textbf{y}$ denote the vector of observations for any of the three stages detailed in Section~\ref{subsec:splicedgammagp}, with associated hyperparameters $\boldsymbol\theta_{1}=\kappa$ (Gamma likelihood) or $\boldsymbol\theta_{1}= \xi$ (GP likelihood). As in Section~\ref{subsec:generalframework}, {let $\mathbf{x} = (\boldsymbol{\eta}^T,\mu,\boldsymbol{\beta}^T,\mathbf{f}^T)^T$ be the latent Gaussian random field},  $\boldsymbol\theta_{2}$ be the vector of hyperparameters of any of the latent models detailed in Section~\ref{subsec:latentstructure}, and $\boldsymbol\theta = (\boldsymbol\theta_{1}^T, \boldsymbol\theta_{2}^T)^T$. Then, from~\eqref{eq:hierarchicalrep}, the joint posterior distribution of parameters and hyperparameters for any of the three stages, can be written as
\begin{align}\label{eq:posterior}
	\text{p}(\mathbf{x},\boldsymbol{\theta} \mid \textbf{y}) &\propto \text{p}(\boldsymbol{\theta})\text{p}(\mathbf{x} \mid \boldsymbol{\theta}_2)\prod_{i\in\mathcal{I}}\text{p}(y(i) \mid x(i),\boldsymbol{\theta}_1)\nonumber\\
	&\propto \text{p}(\boldsymbol{\theta}) |\boldsymbol{Q}_{\boldsymbol{\theta}_2}|^{1/2}\exp\left(-\frac{1}{2}(\mathbf{x}-\boldsymbol{\mu}_{\boldsymbol{\theta}_2})^T\boldsymbol{Q}_{\boldsymbol{\theta}_2}(\mathbf{x}-\boldsymbol{\mu}_{\boldsymbol{\theta}_2}) + \sum_{i\in\mathcal{I}}\log\text{p}(y(i) \mid x(i),\boldsymbol{\theta}_1) \right),
\end{align}
where $\boldsymbol{\mu}_{\boldsymbol{\theta}_2}$ and $\boldsymbol{Q}_{\boldsymbol{\theta}_2}$ are the mean and precision matrix of $\mathbf{x}$, respectively. {Note that for the off-site latent model, {$|\mathcal{I}| = T$} {(number of time points)} for Stages 1 and 2, and $|\mathcal{I}|$ is the number of exceedances in Stage 3, whereas for the SPDE latent model, {$|\mathcal{I}| = ST$} {(number of space-time points, {where $S$ corresponds to the number of stations})} for Stages 1 and 2, whereas $|\mathcal{I}|$ is the number of space-time exceedances in Stage 3}. {We emphasize that each stage is fitted separately, and therefore~\eqref{eq:posterior} applies for each stage independently}. The main objectives of the statistical inference are to extract from~\eqref{eq:posterior} the marginal posterior distributions for each of the elements of the linear predictor vector, and for each element of the hyperparameter vector, i.e.,
	\begin{align}\label{eq:posteriorparam}
	\text{p}(x(i) \mid \boldsymbol{y}) &= \int \text{p}(\boldsymbol{\theta} \mid \boldsymbol{y})\text{p}(x(i) \mid \boldsymbol{\theta}, \boldsymbol{y})\text{d}\boldsymbol{\theta},\qquad \text{p}(\theta_k\mid\boldsymbol{y}) = \int \text{p}(\boldsymbol{\theta}\mid\boldsymbol{y})\text{d}\boldsymbol{\theta}_{-k},
	\end{align}
from which predictive distributions may be derived.

In a Bayesian framework, model estimation is typically performed using simulation-based techniques, such as MCMC methods. Alternatively, approximate methods can be used to cope with the computation of high{-}dimensional integrals needed to obtain posterior distributions. One of such approaches that has become increasingly popular in the last decade is the integrated nested Laplace approximation (INLA;~\citeauthor{rue2009approximate}, \citeyear{rue2009approximate}), where posterior distributions of interest are numerically approximated using the Laplace approximation, therefore avoiding the usually complex updating schemes, long running times, and diagnostic convergence checks of simulation-based MCMC. INLA is designed for latent Gaussian models and therefore, it can be successfully used in a wide variety of applications~(see for instance~\citeauthor{riebler2012estimation}, \citeyear{riebler2012estimation}; \citeauthor{lombardo2017point}, \citeyear{lombardo2017point}; \citeauthor{book126}, \citeyear{book126}). The R-INLA package~\citep{bivand2015spatial} is a convenient interface to the INLA methodology. A wide variety of models are already implemented, and the package is continuously maintained and updated. In particular, our work motivated the INLA implementation of the Gamma and Weibull quantile regressions, both now available to the users. Details regarding the numerical approximations to \eqref{eq:posteriorparam} given by INLA are provided in Section 1 of the Supplementary Material.

%-------------------------------------------------------------------------------
%	FORECAST RESULTS
%-------------------------------------------------------------------------------
\section{Wind speed probabilistic forecasting results}\label{sec:speedresults}
\subsection{Automatic off-site predictor selection based on wind direction}\label{subsec:neighborhoodselection}
An important step to fit our off-site latent model described in Section~\ref{subsec:offsitemodel} is to select a suitable set of off-site predictors $N_\ss$, $\ss\in\mathcal{S}$. Here, we develop a data-driven approach for identifying dominant wind directions, which we then use subsequently to automatically choose the off-site predictors. {This procedure is similar in spirit to that in~\cite{kazor2015role} who identify wind regimes by fitting a Gaussian mixture model to the wind vector.}

\begin{figure}[t!]
\centering
\includegraphics[width=55mm]{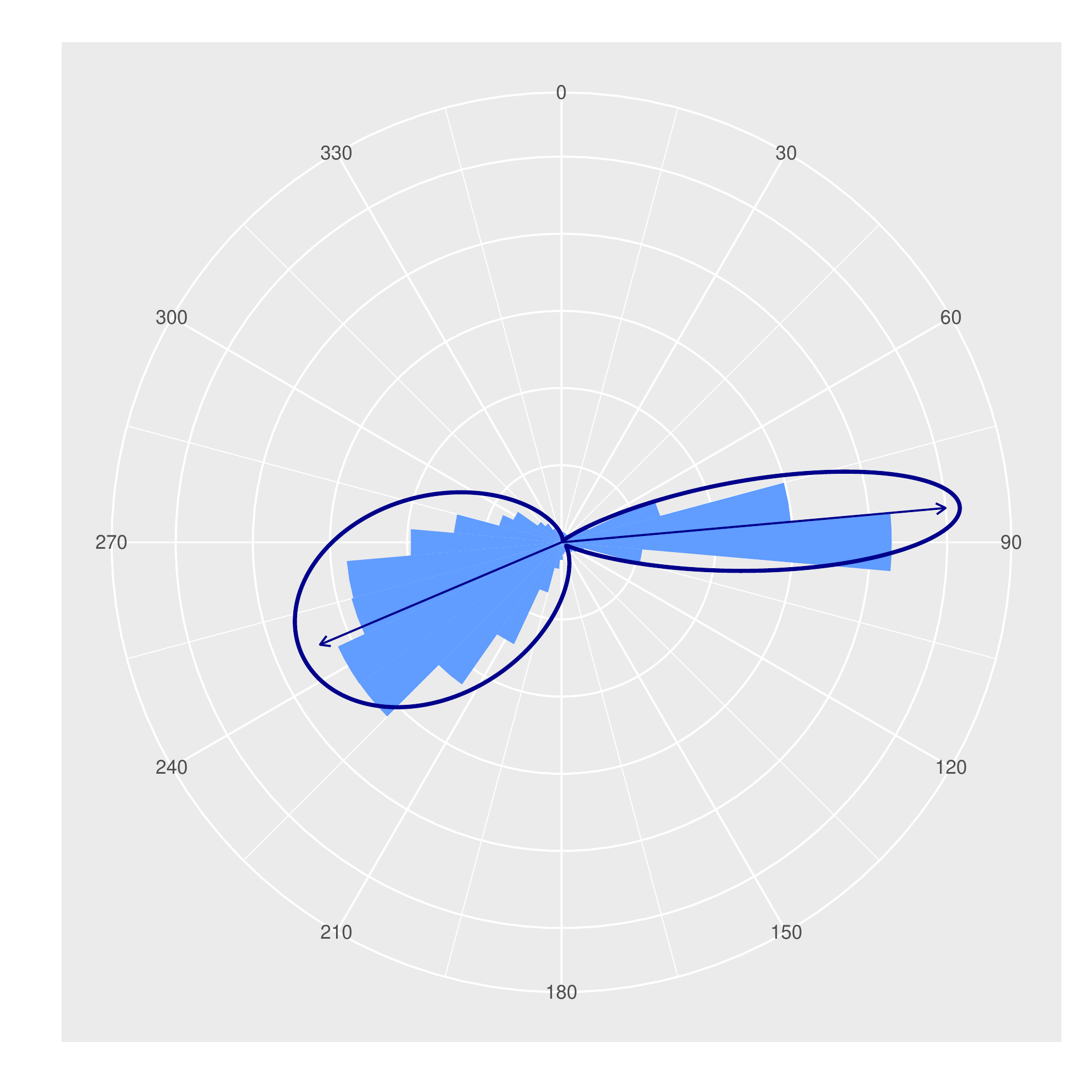}
\includegraphics[width=55mm]{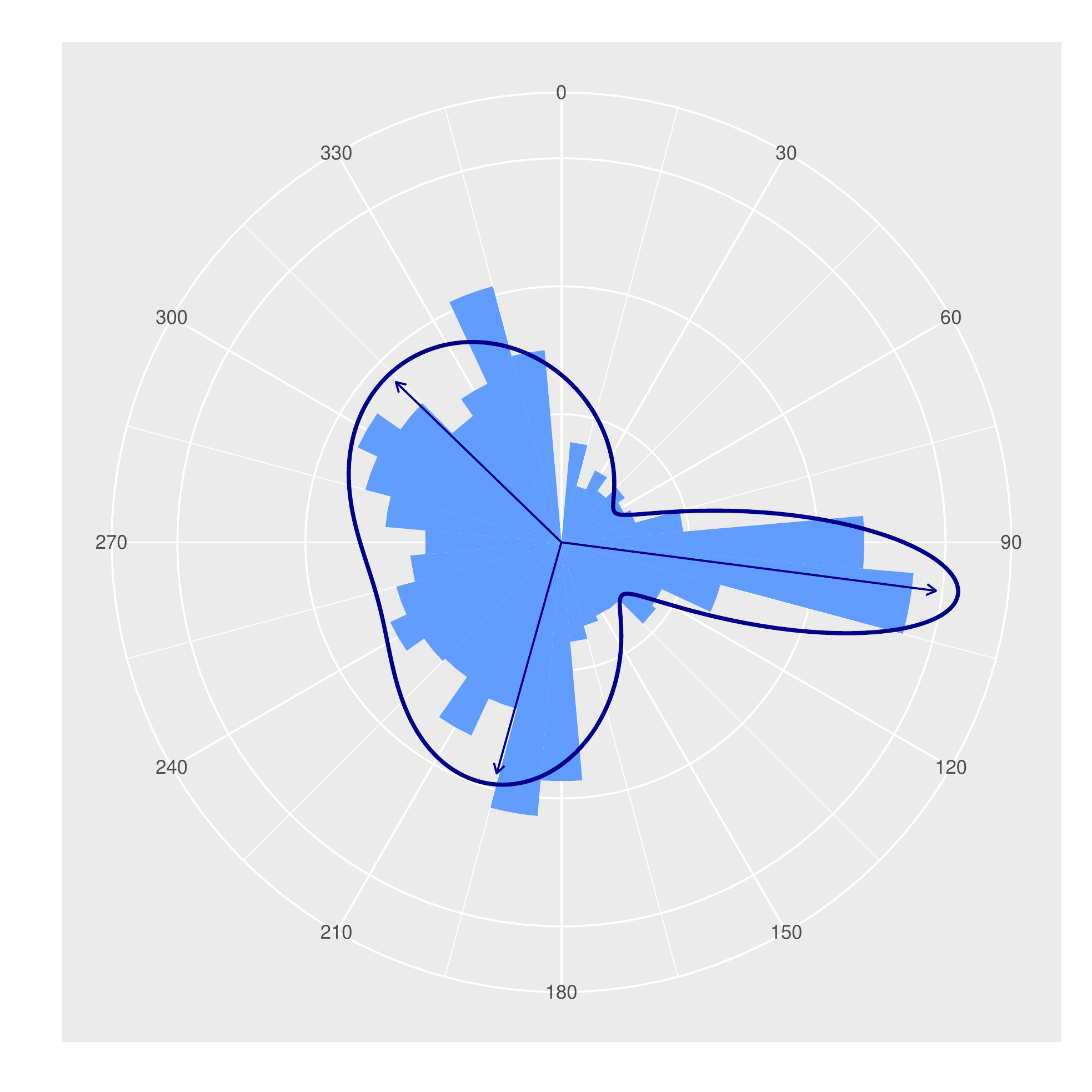}
\includegraphics[width=55mm]{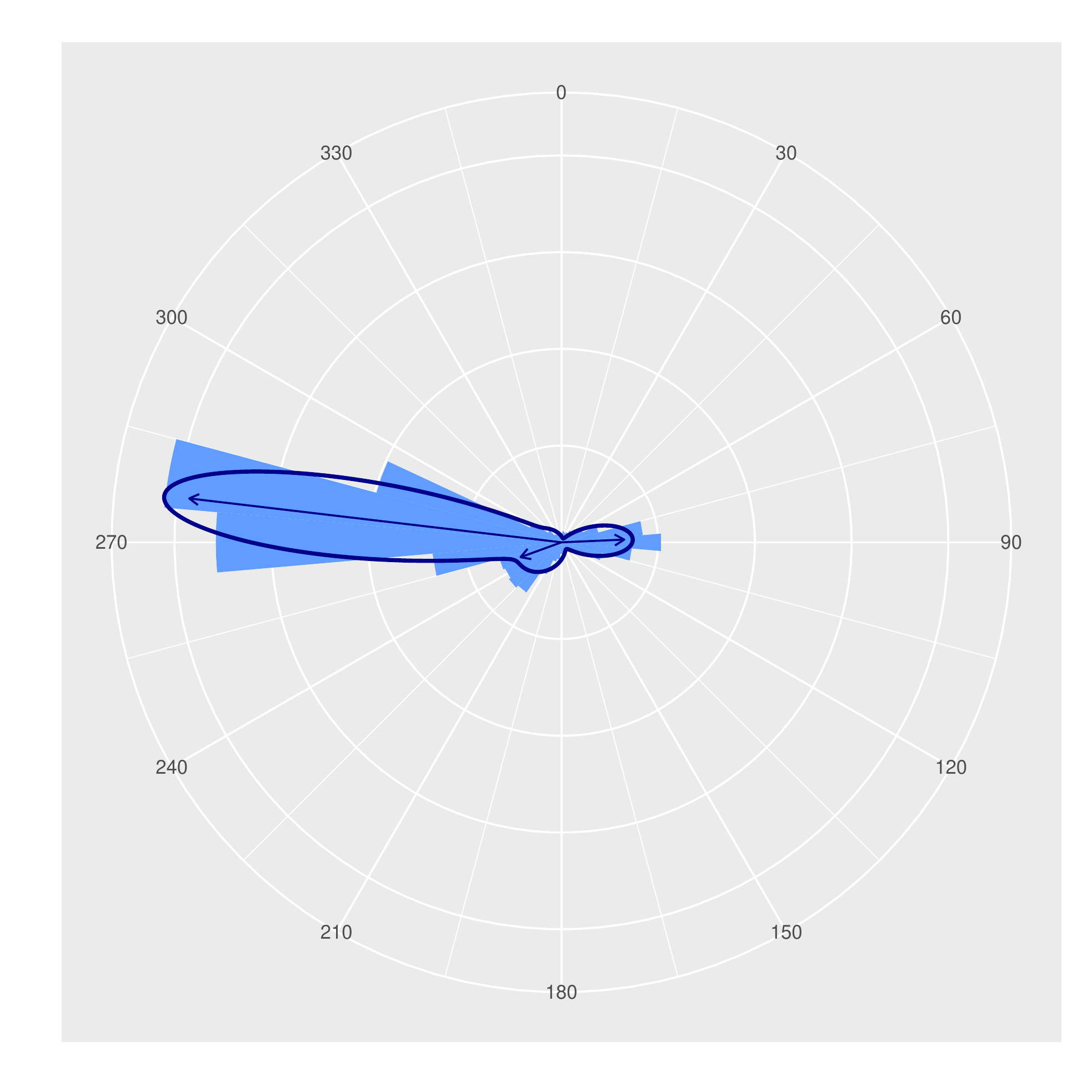}
\caption{\footnotesize Circular histograms and fitted von Mises distribution mixtures with two and three dominating wind directions for Biddle Buttle (BID), Megler (MEG), and Wasco (WAS).}
\label{fig:dominantwind.pdf}
\end{figure}

For each station $\ss\in\mathcal{S}$, we fit a mixture of von Mises circular distributions to the time series of wind directions, $\theta\in[0,2\pi)$. The von Mises density with location parameter $\mu\in\mathbb{R}$ and concentration parameter $\upsilon>0$ is given by $f(\theta\mid\mu, \upsilon) = [\exp\{\upsilon\cos(\theta-\mu)\}]/\{2\pi\mathbf{I}_0(\upsilon)\}$, where $\mathbf{I}_0(\upsilon)$ is the modified Bessel function of order 0. For a mixture of {$M$} von Mises distributions, we identify the dominant wind directions with the locations parameters {$\mu_{\ss,m}$, $m=1,\dots,M$}, and construct the sets of angles {$\mathcal{R}_{\ss,m}(\alpha) = \{\theta\in[0,2\pi):\mu_{\ss,m} - \alpha \leq \theta \leq\mu_{\ss,m} + \alpha\}$}, {for some} $\alpha\in[0,\pi/4]$, defining station-specific directions of influence. Then, the set of off-site predictors for the station $\ss\in\mathcal S$ is chosen as all the stations whose angle with $\ss$ lies in {$\cup_{m=1}^M\mathcal{R}_{\ss,m}(\alpha)$}, such that their distance to $\ss$ is less than some maximum distance $\text{d}_{\text{max}}$. We selected the number of components for the mixture of von Mises distributions via the Bayesian Information Criterion (BIC), also guided by the wind roses displayed in Figure~1 of the Supplementary Material. The angle $\alpha$ depends on the geographical conditions and the distance between stations. We choose $\alpha=\pi/8$ for all stations. We take $\text{d}_{\text{max}}=176$km, which corresponds to the mean distance among all stations. To illustrate our approach, Figure~\ref{fig:dominantwind.pdf} displays the fitted dominant wind directions for Biddle Buttle (BID), Megler (MEG), and Wasco (WAS) stations. The number of off-site predictors for all the stations varies between 2 and 10, while the distance between off-site predictor locations and $\ss$ ranges between 13.3km and 175.3km, for all $\ss\in\mathcal{S}$. {Because the procedure described above is fast, it would be possible to extend it in order to dynamically select potentially different sets of suitable off-site predictors over time.}

% ===========================================================
% Subsection
% ===========================================================
\subsection{Posterior predictive distributions}\label{subsec:posteriorpredictivedist}
Here we briefly explain how to obtain posterior predictive distributions for the 1-hour, 2-hour, and 3-hour ahead probabilistic forecasts of hourly wind speeds, produced by our {spliced Gamma-GP} hierarchical Bayesian model, using the linear predictors introduced in Section~\ref{subsec:latentstructure}. For the SPDE latent  model, we use a rolling training period of length 5 days, whereas for the off-site latent model, we multiply this period by the number of stations in each side of the Cascade Mountains, as a way to balance the effective sample sizes of the SPDE and the off-site latent models. We generate $10,000$ samples from the posterior predictive distribution, for each station, each forecasting time horizon, and each latent model, as follows: we extract the posterior means of the linear predictor and hyperparameters for each stage, and use~\eqref{eq:link} to obtain $10,000$ samples for the Gamma, Bernoulli, and GP predictive distributions. We replace Gamma samples by threshold exceedances (GP samples) whenever the threshold is exceeded, i.e., whenever the associated Bernoulli sample is equal to 1. In other words, the tail of the Gamma distribution is \emph{corrected} by the GP distribution in the presence of exceedances. {Note that a more realistic sampling scheme would be to sample non-exceedances from a truncated Gamma distribution, which implies an additional step between Stages 1 and 2, where a truncated Gamma distribution is fitted up to the threshold $\psi_{\ss,\alpha}(t)$. In this case, all values above the threshold would come strictly from the generalized Pareto distribution.}

{Estimates for the GP shape parameter $\xi$ are obtained by fitting the GP distribution to each training set. The sequence of estimates computed using all the training data indicate that the wind speed distribution exhibits a moderately heavy upper tail with $\hat\xi\approx0.07$--$0.1$. Since we restrict {ourselves} to $\xi\geq0$, our model cannot produce predictive distributions with bounded tails}, {but this {does not} seem to be a major concern {here}; see Figure 7 in Section 5 of the Supplementary Material.}

%-------------------------------------------------------------------------------
% SUBSECTION
%-------------------------------------------------------------------------------
\subsection{Forecast evaluation for extreme and non-extreme wind speeds}\label{subsec:evalforecast} 
We consider performance measures that describe the ability of our {spliced Gamma-GP} model to forecast extreme and non-extreme wind speeds.  {To measure of the overall forecasting ability}, we consider the continuous ranked probability score (CRPS) introduced by~\cite{gneiting2007probabilistic}. The CRPS is a proper scoring rule that quantifies the calibration and sharpness of the forecast~\citep{gneiting2007probabilistic}. {In the following, we describe the CRPS using the notation for the off-site model, but the same definition applies {to} the SPDE model with the corresponding change of notation. The CRPS} {is a function of the observed wind speed $y_\ss(t)$ and the corresponding {predictive} distribution {$\hat{F}_\ss(t)$, $t\in\mathcal{T}$, and it is defined as}
\begin{align}\label{eq:crps}
	\text{CRPS}\{\hat{F}_\ss(t),y_\ss(t)\} &= \int_{-\infty}^\infty [\hat{F}_\ss(t)(x)-\mathds{1}\{y_\ss(t)\leq x\}]^2\text{d}x
\end{align}
The average CRPS for the $h$-hour ahead forecast is 
\begin{equation}\label{eq:acrps}
	\overline{\text{CRPS}}(h)=\frac{1}{T}\sum_{t=1}^{T}\text{CRPS}\{\hat{F}_\ss(t+h),y_\ss(t+h)\}.
\end{equation}

To assess the predictive performance in the upper tail of the distribution, we use the quantile loss {(QL)} function and the threshold-weighted continuous ranked probability score (twCRPS)~\citep{gneiting2011comparing}. The quantile loss function measures the performance of a model to estimate a specific quantile $\tau\in(0,1)$, and it is defined as
\begin{equation*}
	\text{{Q}L}_\tau(y, q) = \begin{cases} 
      \tau(y-q), & \text{if } ,y \geq q, \\
      -(1-\tau)(y-q), & \text{if } , y < q.\\
   \end{cases}
\end{equation*}

{If $Y\sim F$ then $\text{arg}\min_q \E\{\text{{Q}L}_\tau(Y, q)\} = F^{-1}(\tau)$, so this loss function has been used extensively in non-parametric quantile regression; see~\cite{K05}.}

The twCRPS is defined as
\begin{align*}\label{eq:twcrps}
	\text{twCRPS}\{\hat{F}_\ss(t),y_\ss(t)\} &= \int_{-\infty}^\infty [\hat{F}_\ss(t)(x)-\mathds{1}\{y_\ss(t)\leq x\}]^2\omega(x)\text{d}x,
\end{align*}
where $\omega(x)$ is a non-negative weight function on the real line. For $\omega(x)\equiv1$, the twCRPS reduces to the CRPS in~\eqref{eq:crps}.  {We select two different weight functions that highlight our interest {in} the right tail. We set $\omega_1(x)=\mathds{1}\{x\geq r\}$, with $r$ {equal to} the 95\% quantile of the wind speed distribution, and $\omega_2(x) = \Phi(x\mid r, 1)$, where $\Phi(\cdot)$ denotes the standard normal distribution. {While $\omega_2(x)$ yields a proper scoring rule, $\omega_1(x)$ does not because the distribution is left-truncated; see~\cite{lerch2017forecaster}}. The average quantile loss and twCRPS for the $h$-hour ahead forecast is defined as in~\eqref{eq:acrps}. Lower values of these criteria are better.}

  {Since our main goal is to accurately forecast extreme wind speeds, we compare our {proposed spliced Gamma-GP} models against a baseline Gamma model that forecasts wind speeds using only the first stage described in Section~\ref{subsec:splicedgammagp} {(hence, without correcting the upper tail)}. This baseline model assumes the two latent structures detailed in Section~\ref{subsec:latentstructure}. {Given that our proposed model is fairly complex, the} purpose of this comparison is to check if the GP correction of the tail improves the forecast of strong wind speeds.}
 {Table~\ref{table:crps1} shows performance measures for one-hour ahead forecasts for the baseline and the Gamma-GP models at ten selected stations}, {as well as average prediction skills at all stations}. Throughout all the fits, we set the probabilities defined in Section~\ref{subsec:splicedgammagp} as $\alpha = 0.8$ and $\beta = 0.5$.  {The value of $\alpha$ was chosen as a compromise between having a good approximation to the tail of the wind speed distribution and having enough exceedances to fit the GP approximation in each time window. The value of $\beta$ was chosen pragmatically in order to reduce the correlation between the estimated GP parameters {(see Section 6 of the Supplementary Material for further details regarding the GP parametrization)}.}  {{The performance measures are the {average} CRPS, {average} twCRPS using $\omega_1(x)$ and $\omega_2(x)$ as defined before, and {average} quantile loss (QL). {From Table~\ref{table:crps1}, we} can see that the SPDE latent model performs better than the off-site latent model at predicting strong values of wind speeds. The difference might be due to the difficulty of the off-site latent model at estimating the GP shape parameter at each station, while a single shape parameter is assumed in the SPDE latent  model, reducing dramatically the estimated posterior predictive uncertainty by borrowing strength across all stations. Both the off-site and the SPDE latent models appear to be better than their baseline counterparts when focusing on the upper tail of the distribution, showing that the GP correction {may be useful to improve} the forecasting of strong wind speeds, although further diagnostics would be needed to draw firm conclusions. }
%An additional advantage of the SPDE latent  model is that it can be also used to forecast wind speeds at unobserved stations, which cannot be done with the off-site latent model.
%our model is better in all the cases to the persistence forecast, and that for almost all the stations, the correction of the Gamma tail by the GP distribution has a positive effect on the forecasting accuracy.}

\begin{table}[ht]
\centering
 \caption{\footnotesize Performance measures for one-hour ahead forecast using {(left to right)} the off-site model, the off-site baseline model, the SPDE model, and the SPDE baseline model at ten selected stations. {We report the average} continuous ranked probability score (CRPS); {average} threshold-weighted continuous ranked probability score (twCRPS) using {the} indicator weight function $\omega_1(x)$ and {the} normal weight function $\omega_2(x)$; and {average} quantile loss (QL) with $\tau = 0.99$. {The b}ottom line {is} the average for each performance measure over the 20 stations. {The best model is highlighted in bold.}}
\begin{tabular}{l  c  c  c c }
\cline{3-4}
&\multicolumn{3}{c}{\hskip4.3cm twCRPS} \\
%\cline{3-5}
\hline
 Station &CRPS& $\omega_1(x)=\mathds{1}\{x\geq r\}$ & $\omega_2(x) = \Phi(x\mid r, 1)$ & QL ($\tau=0.99$)\\ 
  \hline
AUG & 0.92/0.93/0.92/0.92 & 0.06/0.06/0.06/0.05 & 0.06/0.06/0.06/0.09 & 0.80/0.81/0.76/0.80 \\ 
  BID & 0.90/1.09/0.91/0.92 & 0.09/0.09/0.08/0.10 & 0.09/0.09/0.06/0.11 & 0.77/1.08/0.71/1.06 \\ 
  FOR & 0.39/0.39/0.38/0.39 & 0.04/0.04/0.04/0.06 & 0.03/0.02/0.02/0.07 & 0.38/0.40/0.39/0.39 \\ 
  HOO & 0.61/0.62/0.60/0.61 & 0.04/0.04/0.04/0.07 & 0.05/0.05/0.04/0.09 & 0.56/0.58/0.53/0.61 \\ 
  KEN & 1.14/1.15/1.01/1.13 & 0.09/0.09/0.06/0.10 & 0.05/0.05/0.05/0.08 & 0.95/0.98/0.90/0.10 \\ 
  MEG & 0.67/0.68/0.67/0.67 & 0.07/0.07/0.04/0.08 & 0.07/0.07/0.06/0.10 & 0.62/0.65/0.61/0.82 \\ 
  NAS & 0.69/0.69/0.50/0.75 & 0.06/0.06/0.05/0.09 & 0.06/0.06/0.07/0.09 & 0.65/0.67/0.64/0.85 \\ 
  SML & 0.94/0.95/0.50/0.60 & 0.06/0.06/0.04/0.10 & 0.07/0.07/0.07/0.07 & 0.91/0.93/0.81/0.13 \\ 
  SUN & 1.16/1.17/0.71/1.92 & 0.09/0.09/0.09/0.13 & 0.08/0.08/0.07/0.12 & 1.11/1.14/1.01/1.19 \\ 
  WAS & 0.82/0.82/0.81/0.81 & 0.06/0.05/0.05/0.07 & 0.06/0.06/0.05/0.06 & 0.77/0.79/0.70/0.81 \\ 
  \hline
  Avg.& 0.82/0.85/{\bf 0.75}/0.84 & 0.07/0.07/{\bf 0.06}/0.08 & 0.07/0.07/{\bf 0.06}/0.08 & 0.76/0.79/{\bf 0.73}/0.78  \\
\hline
 \end{tabular}
\label{table:crps1}
\end{table}

We assess the calibration of our probabilistic forecasts using reliability diagrams (see, e.g., \citeauthor{lenzi2017spatial}, \citeyear{lenzi2017spatial}). Reliability refers to the ability of the model to match the observation frequencies. The diagram is constructed as follows: for every station, we compute the nominal coverage rate, which is the proportion of times that the cumulative distribution of our {spliced Gamma-GP} model is below a certain threshold. Our model is well calibrated if this proportion is close to the observed frequencies. Figure~\ref{fig:reliability.pdf} shows reliability diagrams for the off-site latent model ({coral} line) and the SPDE latent  model ({cyan} line), {as well {as for} their baseline counterparts (in green and purple, respectively)}. We can see that both models tend to overestimate the wind speed quantiles smaller than the median. The off-site latent model underestimates the wind speed values larger than the median, whereas the SPDE latent  model is better calibrated at higher quantiles.

\begin{figure}[t!]
\centering
	\includegraphics[width=65mm]{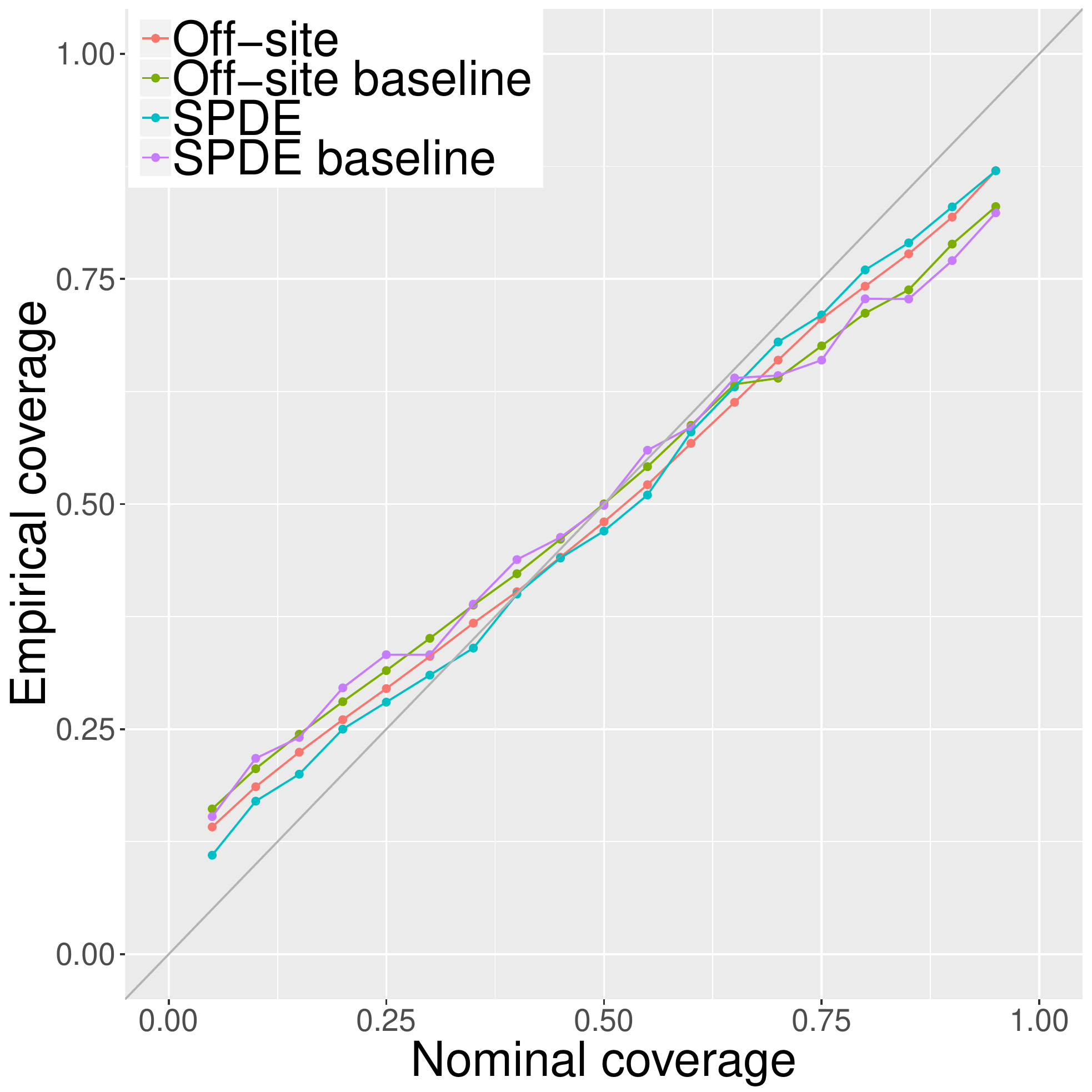}
\caption{\footnotesize Reliability diagram for the off-site model (coral line), off-site baseline model (green line),  SPDE model (cyan line), and SPDE baseline model (purple line).}
\label{fig:reliability.pdf}
\end{figure}

 {Finally, to compare the ability of our proposed models to forecast extreme wind speeds, we compute the {pseudo-uniform scores {$\hat{u}_{\ss,k}(t+h) = \hat{F}_{\ss,k}(t+h)(y_\ss(t+h))$ for all forecasted values $y_\ss(t+h)$, $h=1,2,3$, where $\hat{F}_{\ss,k}(t+h)$ is the predictive distribution for location $\ss\in\mathcal{S}$ and time $(t+h)$ based on the $k$-th training set. Note that we here use the notation for the off-site model, but the same definition applies to the SPDE model with the corresponding change of notation}. We then plot the histogram of {$\{\hat{u}_{\ss,k}(t+h)\}_{k\geq1}$} conditional on being greater than 0.6. We refer to such diagnostics as conditional Probability Integral Transform (PIT) plots. The results {for 1h ahead forecasts} are shown in Figure~\ref{fig:pit.pdf}. Because we condition on {$\hat{u}_{\ss,k}(t+h)>0.6$}, the conditional PIT plots are informative about the model ability to forecast moderately strong wind speeds using the Gamma model described in Stage 1, and strong wind speeds using the GP model in Stage 3; recall Section~\ref{subsec:splicedgammagp}.} We can see that both the off-site model and the SPDE model tend to outperform their baseline {counterparts}, while the SPDE model seems to perform {slightly} better than the off-site model {at} forecasting strong wind speeds. }

\begin{figure}[H]
\centering
\includegraphics[width=39mm]{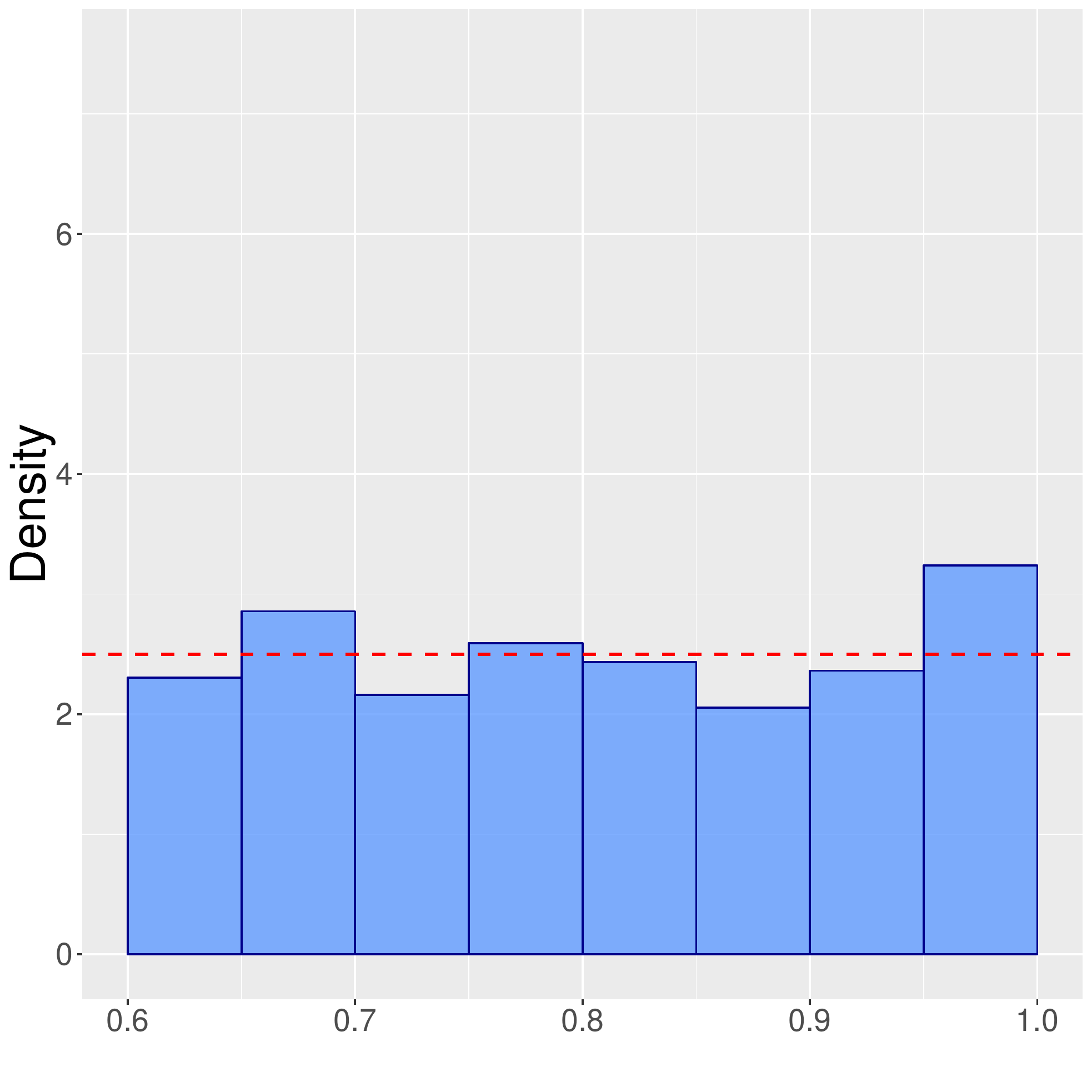}
\includegraphics[width=39mm]{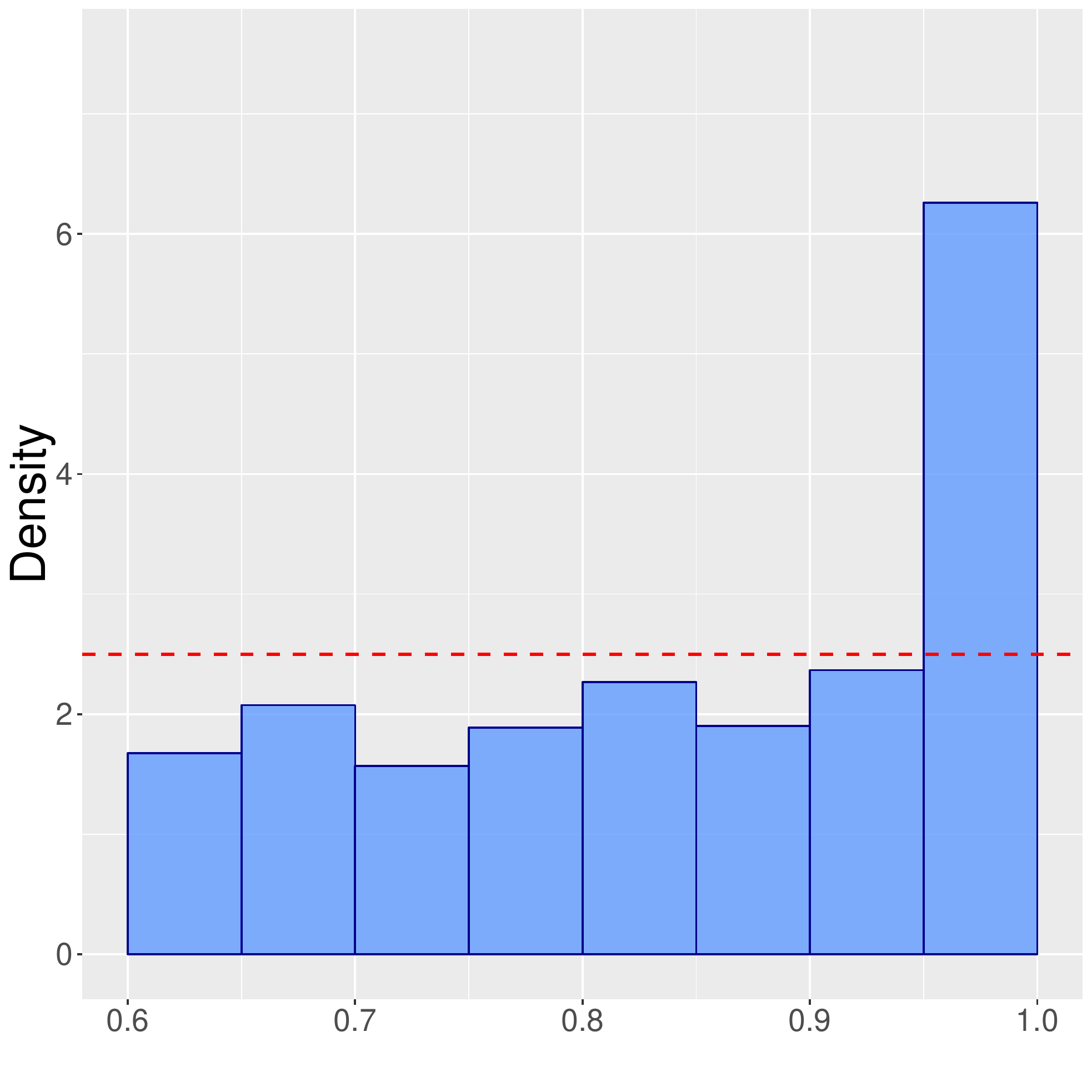}
\includegraphics[width=39mm]{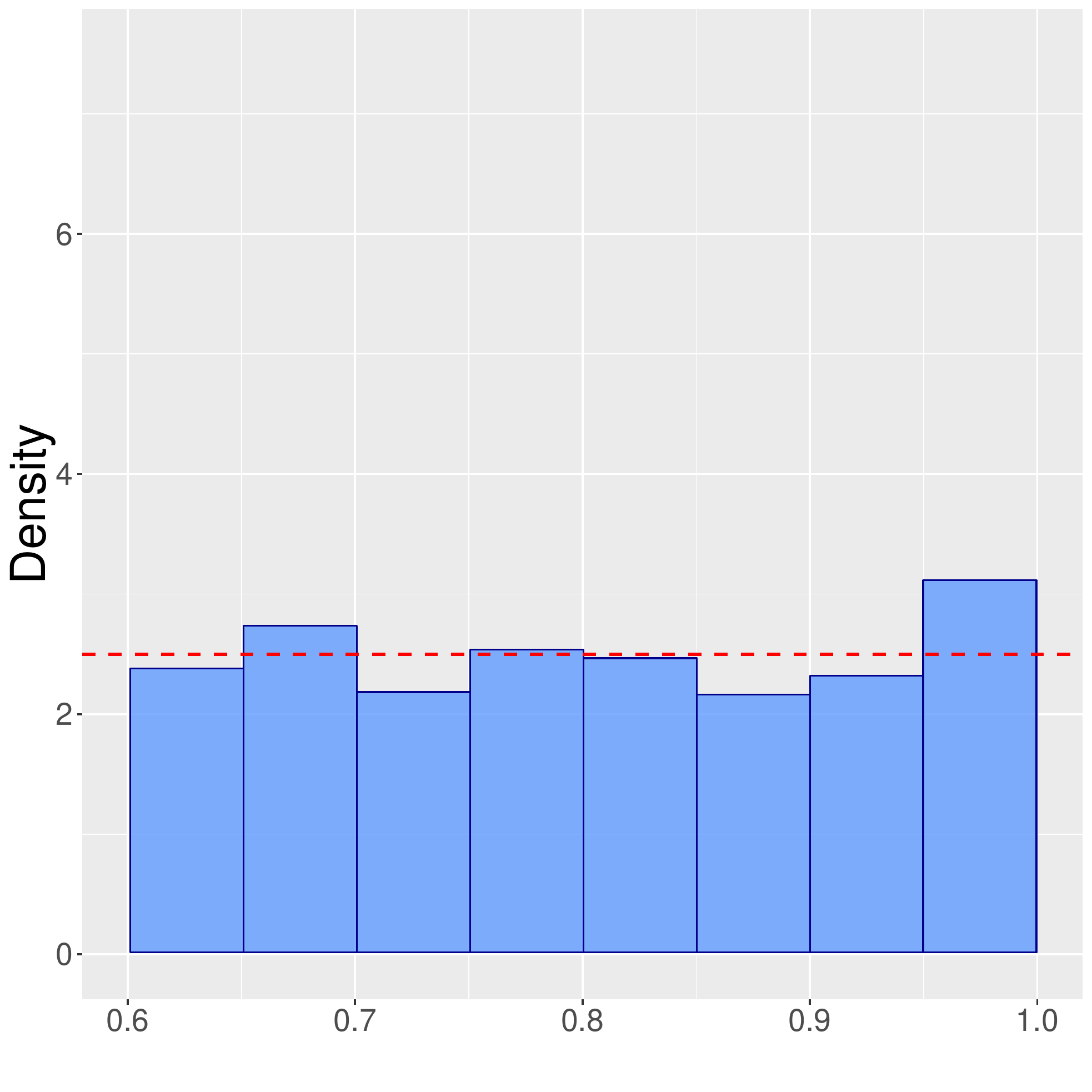}
\includegraphics[width=39mm]{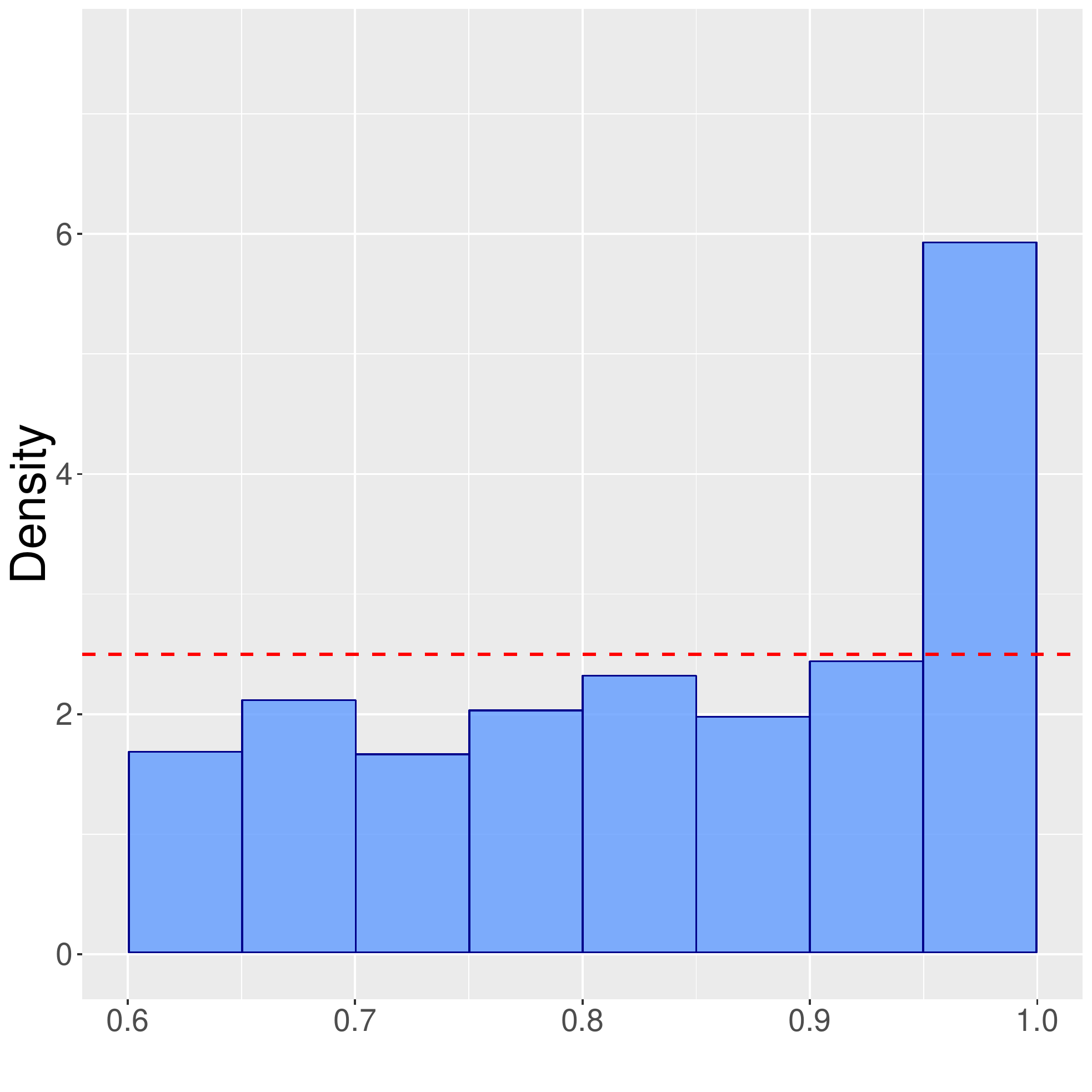}\\
\includegraphics[width=39mm]{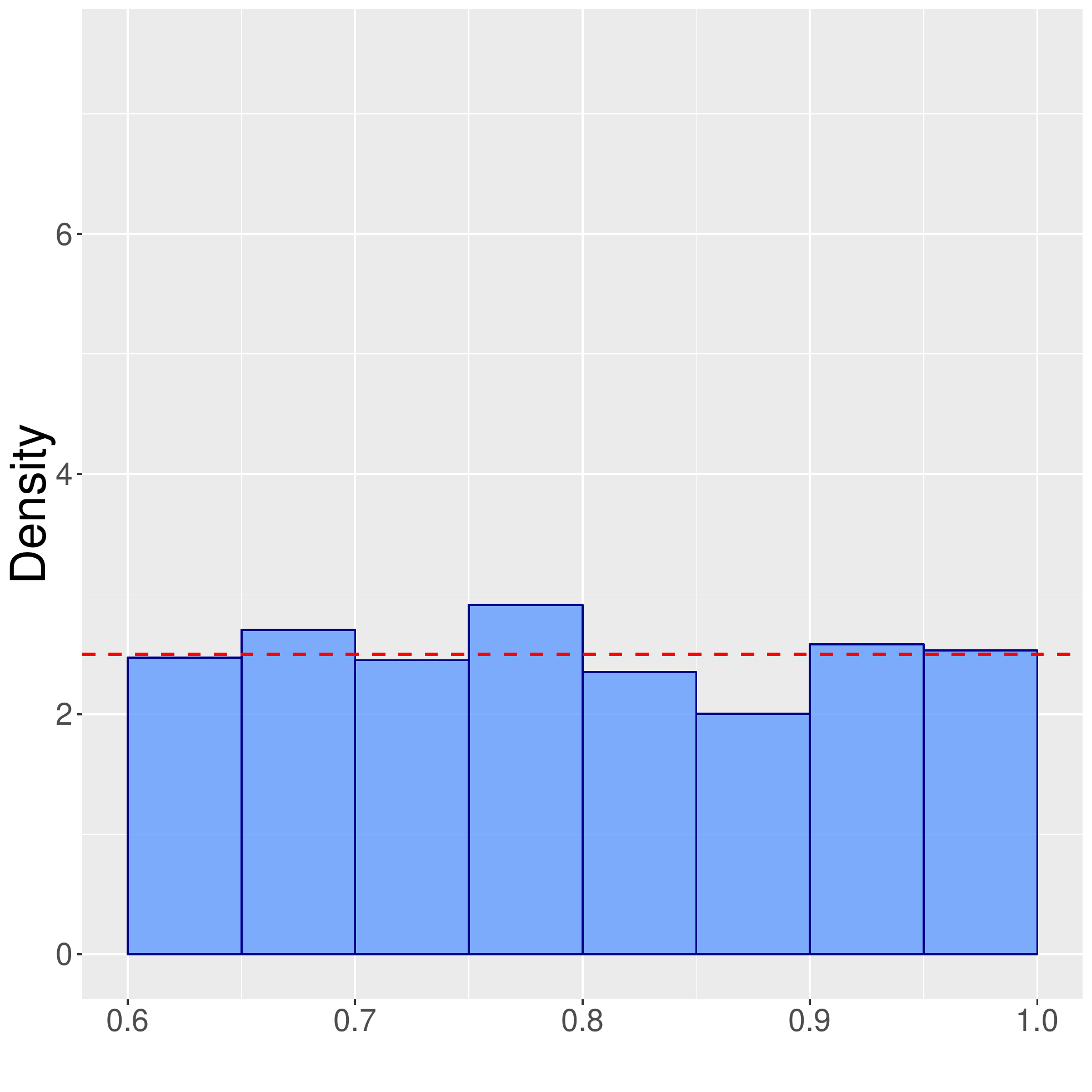}
\includegraphics[width=39mm]{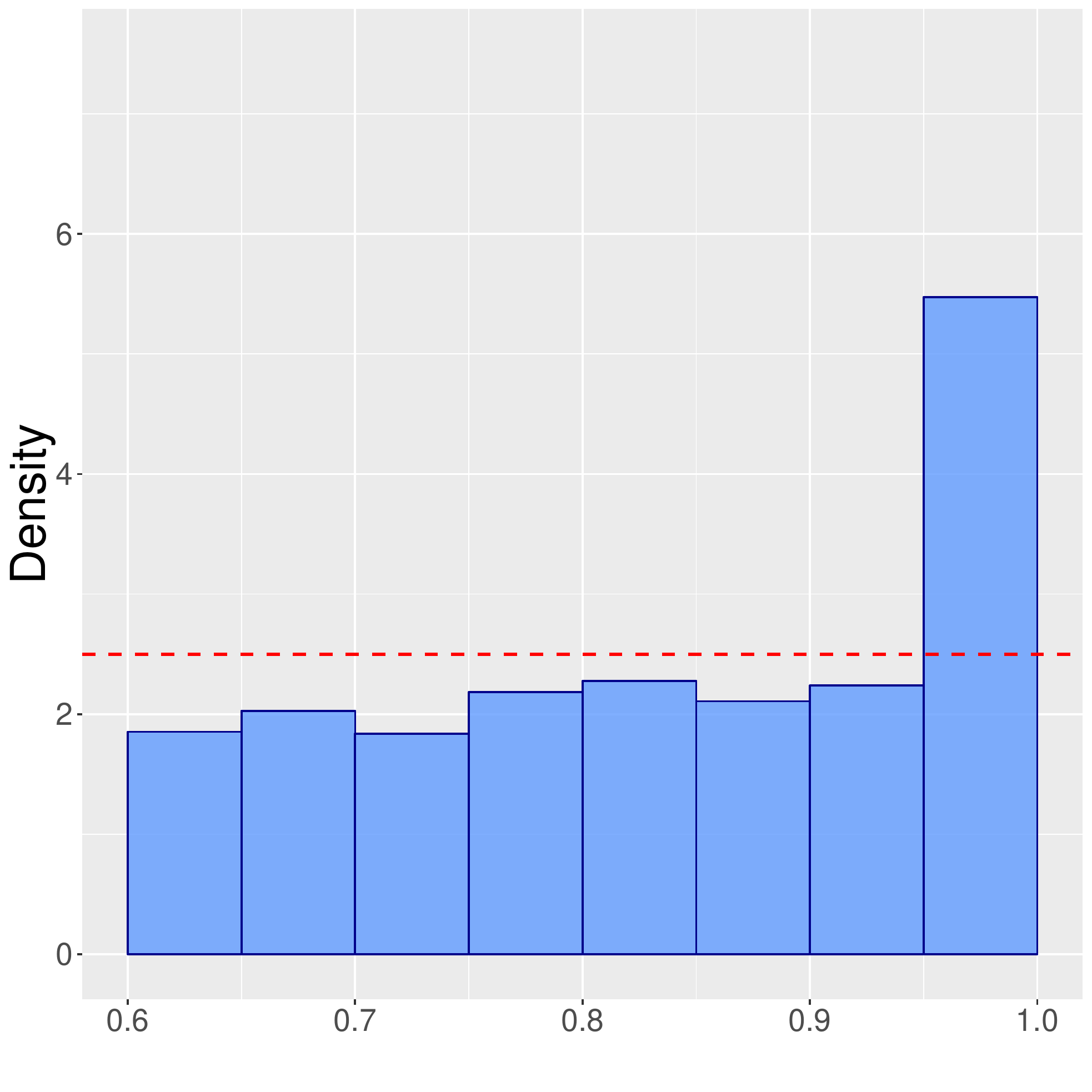}
\includegraphics[width=39mm]{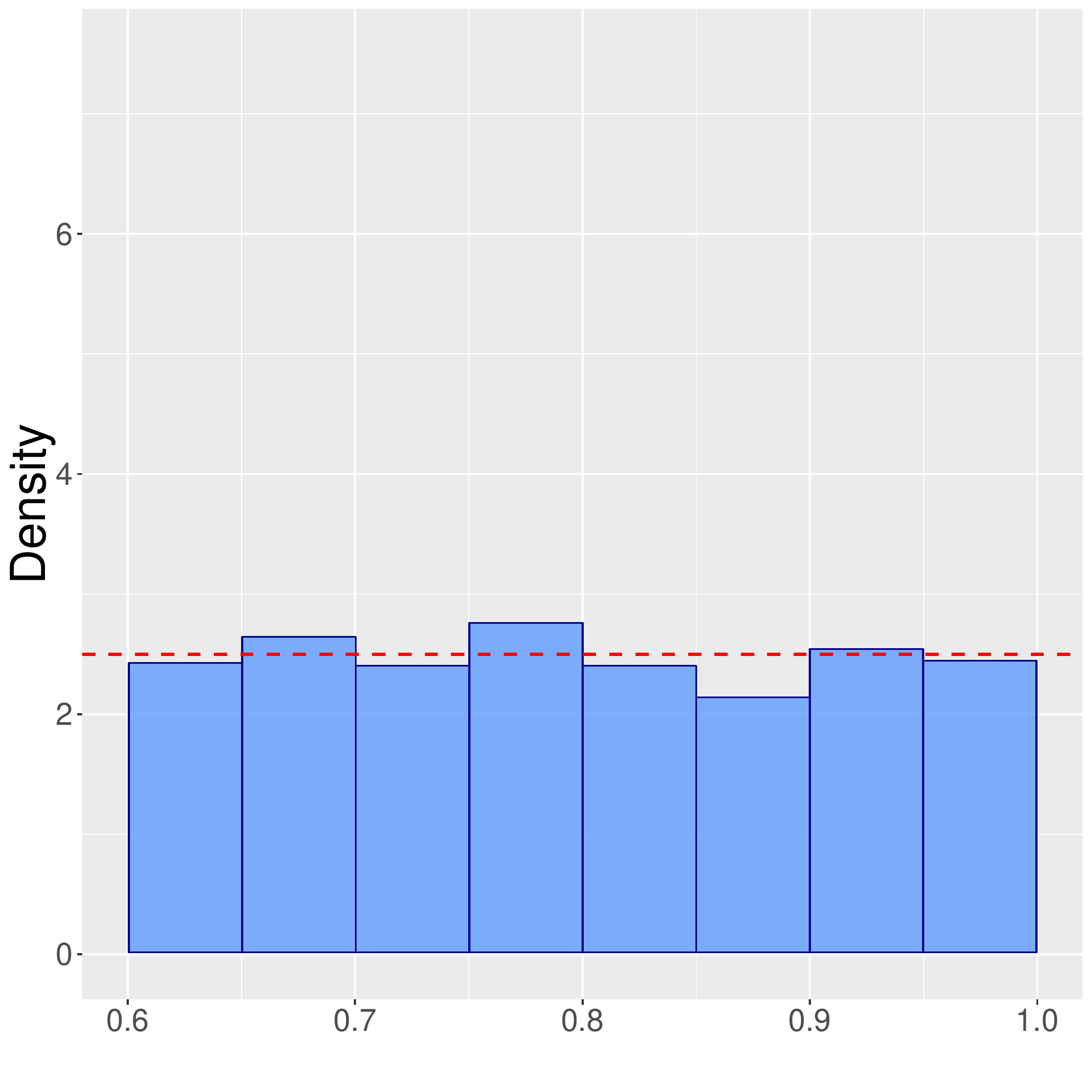}
\includegraphics[width=39mm]{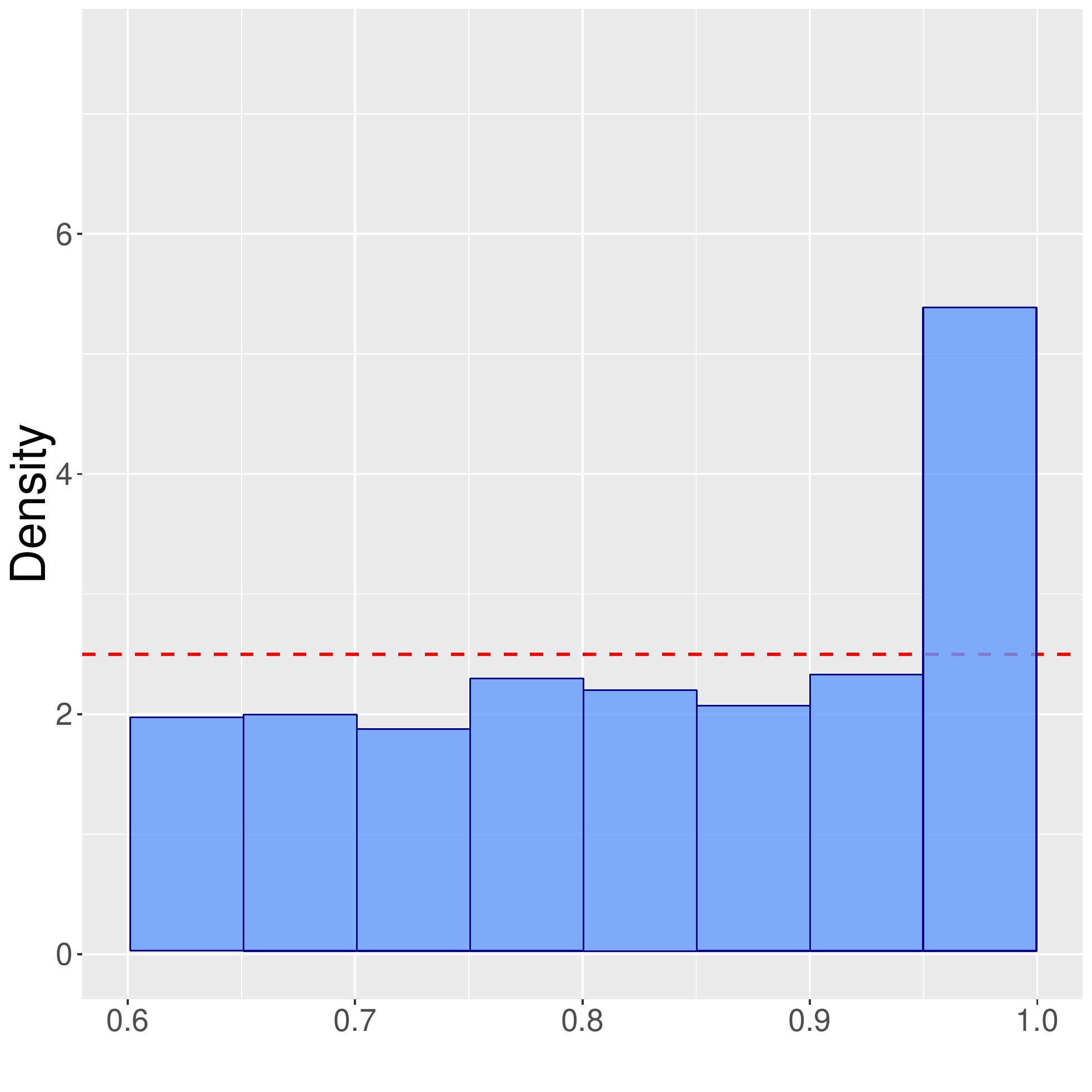}\\
\includegraphics[width=39mm]{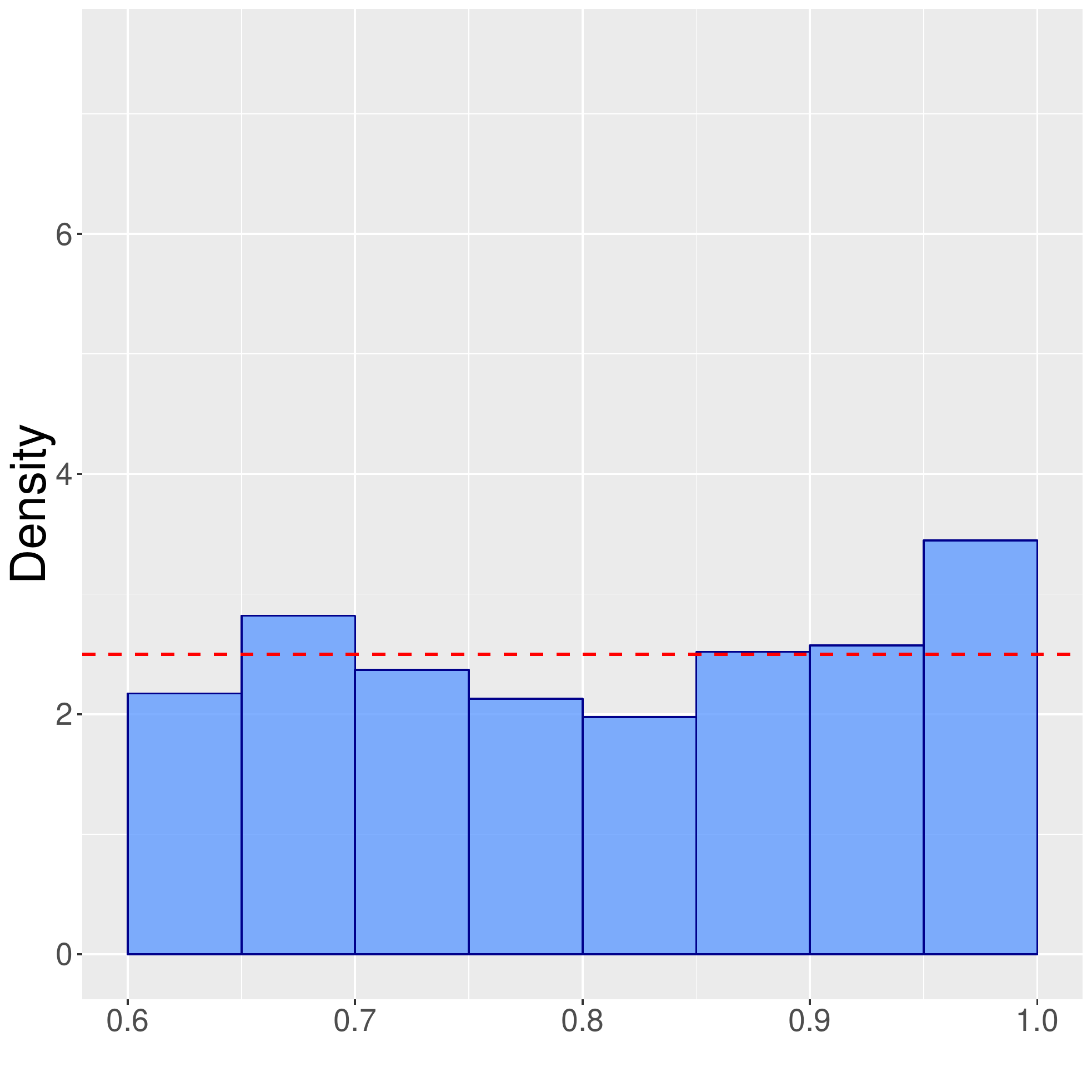}
\includegraphics[width=39mm]{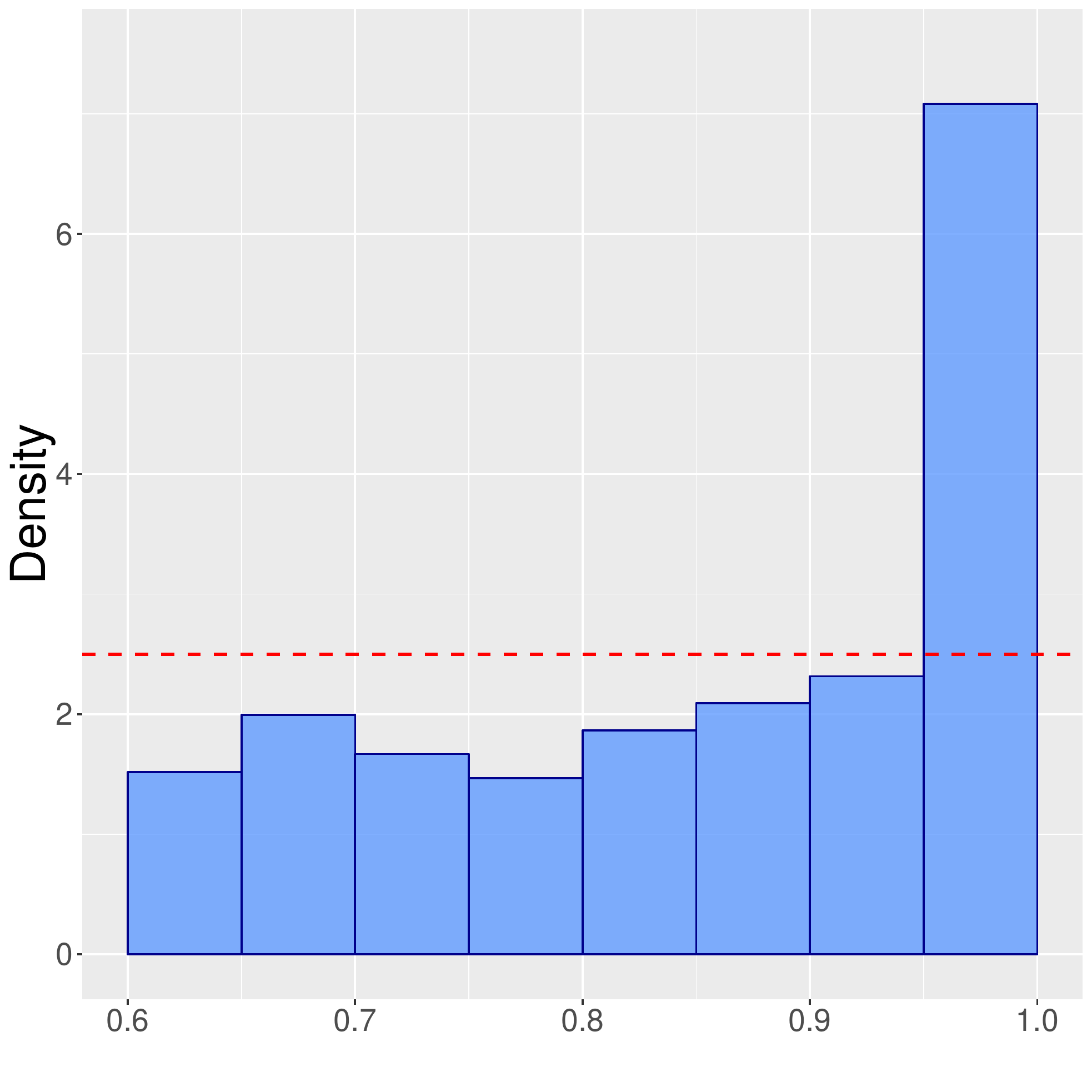}
\includegraphics[width=39mm]{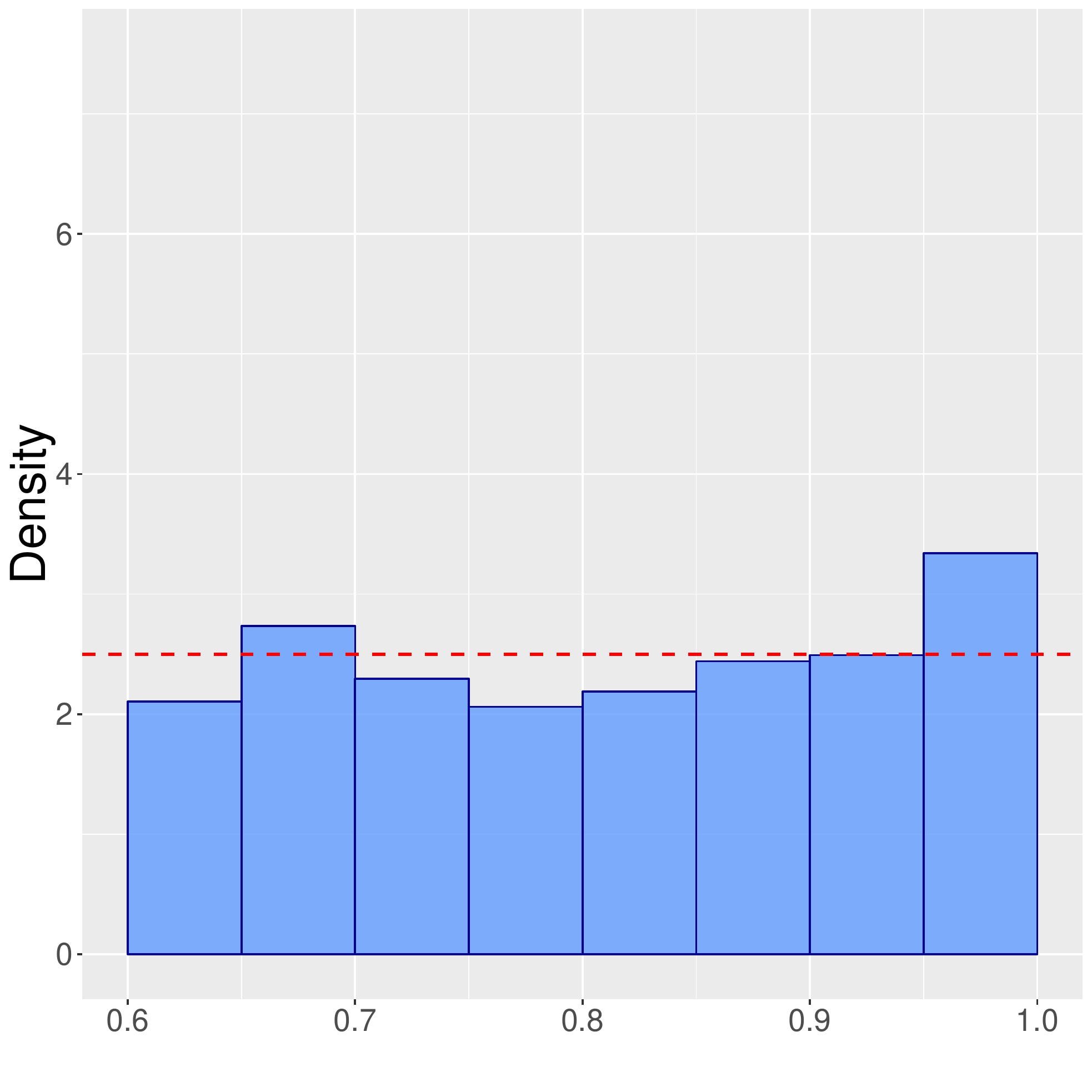}
\includegraphics[width=39mm]{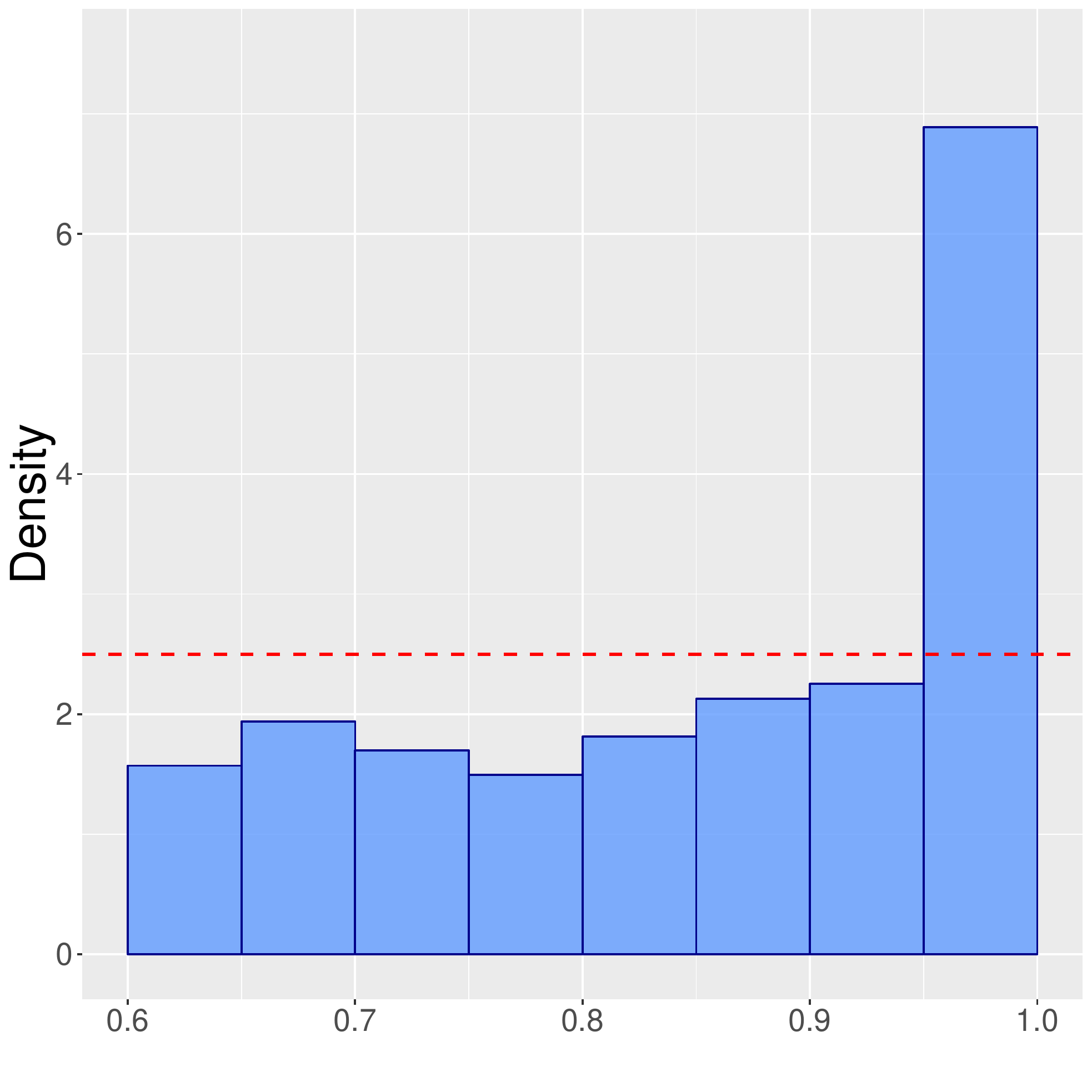}
%\vskip-0.2cm
\caption{\footnotesize {Conditional} PIT {plots} {for 1h ahead forecasts} for the off-site model (first column), the off-site baseline model (second column), the SPDE model (third column), and the SPDE baseline model (fourth column) at three selected stations {(different rows). Uniform histograms imply good forecasting ability}.}
\label{fig:pit.pdf}
\end{figure}

\begin{textblock}{0.5}(1.2,2.3)
  \rotatebox{90}{\bf \black{BID}}
\end{textblock}

\begin{textblock}{0.5}(1.2,4.4)
  \rotatebox{90}{\bf \black{MEG}}
\end{textblock}

\begin{textblock}{0.5}(1.2,6.7)
  \rotatebox{90}{\bf \black{WAS}}
\end{textblock} 

 %-------------------------------------------------------------------------------
%	CONCLUSIONS
%-------------------------------------------------------------------------------

\section {Conclusion}\label{sec:conclusion}
In this work, we {have} explored a {hierarchical} Bayesian  {spliced Gamma-GP} model designed to forecast extreme wind speeds. Our model corrects the tail of the Gamma distribution by a generalized Pareto distribution in the presence of exceedances. Each stage of our model belongs to the class of latent Gaussian models, for which the integrated nested Laplace approximation (INLA;~\citeauthor{rue2009approximate}, \citeyear{rue2009approximate}) method is well-suited. Considering an additive {latent} structure, we proposed two types of linear predictors describing the spatio-temporal dynamics of wind speeds. The first linear predictor includes off-site information in terms of lagged wind speeds from neighboring stations. To select the neighbors, we propose an automatic method to identify dominant wind direction patterns based on {mixtures} of von Mises circular distributions. The second linear predictor considers a spatio-temporal term with Mat\'ern covariance structure {(driven by a stochastic partial differential equation)}, that varies in time according to a first-order autoregressive dynamics. In terms of forecasting extreme wind speeds, both models seem to perform decently well, although the SPDE latent  model is better calibrated at high quantiles, because it better exploits spatial information. It would be interesting to explore the potential of the SPDE latent  model to predict extreme wind speeds at unobserved locations, which could be helpful for optimal design of wind farms.

Thanks to the very powerful and fast INLA estimation approach, we can implement in a reasonable amount of time complex hierarchical spatial models that are well suited to wind speed data. {Specifically, each set of {1h--3h} ahead forecasts using a single core took less than 2 minutes {for} the off-site model, and less than 20 minutes {for} the SPDE model. The parallelization of these computations was done using resources for distributed computing.} 

By selecting a suitable distribution in the first stage, our {spliced Gamma-GP} model for exceedances can be easily adapted to model and forecast other types of environmental data. {We fitted the bulk and the tail of the wind speed distribution separately, as extreme events usually behave differently from {low and moderately large} events, and therefore only extreme observations {may} give information about the tail of the distribution~\citep{rootzen2018multivariate}. If the model for the bulk is misspecified, then the $\alpha$-quantile might not be well estimated. But unless the fitted proportion of exceedances deviates considerable from the {truth (i.e., the parameter $\alpha$)}, this should not considerably affect the fit for the tail.}

{Improving the quality of wind power forecasts is a constant challenge, but there are many possible directions on how to incorporate a rigorous extreme value analysis to the estimation of wind speed. For instance, a better representation of the physical phenomena involved in the generation of strong wind speeds may be achieved by introducing outputs from numerical climate models. Alternatively, non-stationary spatial models with local anisotropies informed by wind direction could be developed. Moreover, for a broader description of the tail of the wind speed distribution, it would be interesting to fit the GP distribution with $\xi\in\mathbb{R}$ (i.e., not only $\xi\geq0$), {which could be useful for lighter-tailed data}. Nevertheless, designing suitable shrinkage priors for $\xi<0$ is still an open question. For computational convenience (and because of the constraints with INLA), we have assumed conditional independence of the data given the latent process. In cases where strong (tail) dependence prevails, more specialized extreme-value models should be considered, such as generalized Pareto processes (\citealp{thibaud2015efficient}), max-stable models (\citealp{huser2014space}), and flexible copula models (\citealp{camilo2017local}). Although these models are attractive from a theoretical viewpoint, they are very cumbersome to fit, especially in high dimensions.}}

%-------------------------------------------------------------------------------
%	ACKNOWLEDGEMENTS
%-------------------------------------------------------------------------------

%\section*{Acknowledgements}
%We thank Amanda Hering for helpful discussion and suggestions, and for providing the wind speed data.  We also extend our thanks to Thomas Opitz for helpful discussion, and to Amanda Lenzi for her constructive suggestions regarding the INLA implementation. Support from the KAUST Supercomputing Laboratory and access to Shaheen is also gratefully acknowledged. This publication is based upon work supported by the King Abdullah University of Science and Technology (KAUST) Office of Sponsored Research (OSR) under Award No. OSR-CRG2017-3434.

%-------------------------------------------------------------------------------
%	REFERENCES
%-------------------------------------------------------------------------------
\baselineskip 10pt
\bibliographystyle{CUP}
\bibliography{References.bib}

\end{document}